\documentclass[11pt]{article}

\newtheorem{theorem}{Theorem}[section]
\newtheorem{lem}{Lemma}[section]
\newtheorem{pro}{Proposition}[section]
\newtheorem{cor}{Corollary}[section]
\newtheorem{conj}{Conjecture}[section]

\newtheorem{rem}{Remark}[section]
\newtheorem{com}{Comments}[section]
\newtheorem{ex}{Example}[section]
\newtheorem{defi}{Definition}[section]
\newtheorem{hyp}{Assumption}[section]

\newcommand{\bt}{\begin{theorem}}\newcommand{\et}{\end{theorem}}
\newcommand{\bl}{\begin{lem}}\newcommand{\el}{\end{lem}}
\newcommand{\bp}{\begin{pro}}\newcommand{\ep}{\end{pro}}
\newcommand{\bcor}{\begin{cor}}\newcommand{\ecor}{\end{cor}}
\newcommand{\bconj}{\begin{conj}}\newcommand{\econj}{\end{conj}}

\newcommand{\bd}{\begin{defi} \rm }\newcommand{\ed}{\end{defi} }
\newcommand{\brem }{\begin{rem} \rm }\newcommand{\erem }{\end{rem}}
\newcommand{\bcom}{\begin{com} \rm }\newcommand{\ecom }{\end{com}}
\newcommand{\brems }{\begin{rem} \rm }\newcommand{\erems }{\end{rem}}
\newcommand{\bex}{\begin{ex} \rm }\newcommand{\eex}{\end{ex}}
\newcommand{\bhyp}{\begin{hyp} \rm }\newcommand{\ehyp}{\end{hyp}}

\def\proof{\noindent {\it {\textbf{Proof}}}.$\;\,$}

\newcommand{\dcb}{\begin{array}{lll}}
\newcommand{\dce}{\end{array}}
\newcommand{\ebe}{\begin{enumerate}\setlength{\baselineskip}{13pt}\setlength{\parskip}{5pt}}
\newcommand{\dbe}{\end{enumerate}}

\newcommand{\ibegin}{\begin{itemize}\setlength{\baselineskip}{19pt}\setlength{\parskip}{7pt}}
\newcommand{\iend}{\end{itemize}}

\newcommand{\desb}{\begin{description}}
\newcommand{\dese}{\end{description}}

\usepackage{xspace,cancel}
\usepackage{bold-extra,chicago}
\usepackage{bm, enumerate}
\usepackage{dsfont}
\def\href#1#2{#2}
\usepackage[dvips]{graphicx}
\usepackage[dvips,usenames]{color}
\usepackage[xcolor]{pstricks}
\usepackage{latexsym, amsfonts, amssymb, amsmath}
\usepackage{alltt,latexsym, varioref,keyval, nameref, verbatim}

\newtheorem{exo}{Exercice}
\def\bexo{\begin{exo}\rm}
\def\eexo{\end{exo}}
\def\sp{,\,}

\def\gg{{\mathbb G}}
\def\Q{{\mathbb Q}}
\def\cA{{\cal {A}}}
\def\G{{\cal G}}

\newcommand{\beqa}{\begin{eqnarray}}
\newcommand{\eeqa}{\end{eqnarray}}
\newcommand{\beqe}{\beqa\begin{aligned}}
\newcommand{\eeqe}{\end{aligned}\eeqa}
\def\bea{\begin{eqnarray*}}
\def\eea{\end{eqnarray*}}

\def\Q{{\mathbb Q}}\def\Q{{\mathbb P}}

\def\E{{\mathbb E}}
\def\cC{{\cal C}}

\def\cH_T{{\cal H}}

\def\G{{\F}}

\def\bal{\begin{aligned}}
\def\eal{\end{aligned}}

\def\M{m}

\def\M{m}

\newcommand{\COMM}[1]{}
\renewcommand{\COMM}[1]{\begin{quote}\begin{red} \small \sf #1\end{red}\end{quote}}

\def\M{m}

\newcommand{\bel}{\bea\bal}
\newcommand{\eel}{\end{aligned}\eea}
\def\bal{\begin{aligned}}
\def\eal{\end{aligned}}
\newcommand{\beql}[1]{\beqa\label{#1}\bal}
\newcommand{\eeql}{\eal\eeqa}
\renewcommand{\bel}{\[\bal}
\renewcommand{\eel}{\end{aligned}\]}

\def\bal{\begin{aligned}}
\def\eal{\end{aligned}}

\def\beq{\begin{eqnarray}}
\def\eeq{\end{eqnarray}}

\def\E{{\mathbb E}}

\def\G{{\cal G}}
\def\cH{{\cal H}}

\def\R{\mathbb{R}}

\def\ind{\mathds{1}}

\def\db{d{{B}}_s}

\def\Q{\mathbb{Q}}

\def\tb{\tilde{b}}

\def\bal{\begin{aligned}}
\def\eal{\end{aligned}}

\def\kappa{S^{\star}}

\def\ind {1\!\!1}\def\ind{\mathds{1}}
\def\I{\ind}

\def\qqq{\quad\quad\quad}
\def\index#1{}

\newcommand{\finproof}{\rule{4pt}{6pt}}

\def\qr#1{\eqref{#1}}

\def\cM{\mathcal{M}}
\def\cG{\mathcal{G}}
\def\eee{\end{document}}
 \def\tb{{\bar{\tau}}}

\def\db{\tb}
\def\ttd{{t \in [0,\tb]}}

\newcommand{\indi}[1]{\I_{\{{#1}\}}}
\def\RM{R}\def\RM{\mbox{ES}}
\def\sr#1{Sect.~\ref{#1}}
\def\M{C}\def\M{\mbox{IM}}
\def\lp{loss-and-profit\xspace}\def\lp{loss\xspace}
\def\lps{loss-and-profits\xspace}\def\lps{losses\xspace}
\def\cVb{\mathcal{V}}\def\cVb{\mbox{RC}}\def\cVb{\overline{\mbox{RC}}}\def\cVb{\widehat{\mbox{RC}}}
\def\cV{\tilde{\mathcal{V}}}\def\cV{\widehat{\mbox{RC}}}\def\cV{\mathcal{W}}\def\cV{\mbox{RC}}
\def\Theta{\mbox{TRC}}
\def\lossb{\varrho}\def\lossb{\overline{L}}\def\lossb{\widehat{L}}
\def\loss{\tilde{\varrho}}\def\loss{\hat{L}}\def\loss{L}

\def\TKVA{KVA\xspace}

\newcommand{\VaR}{\text{VaR}}

\begin{document}
\title{Capital Valuation Adjustment and Funding Valuation Adjustment}
\author{{Claudio Albanese\textsuperscript{1,2},
Simone Caenazzo\textsuperscript{1} and St\'ephane Cr\'epey\textsuperscript{3} }}

{\let\thefootnote\relax\footnotetext{\textsuperscript{1} \textit{Global Valuation Ltd, London.}}}
{\let\thefootnote\relax\footnotetext{\textsuperscript{2} \textit{CASS School of Business, London.}}}

{\let\thefootnote\relax\footnotetext{\textsuperscript{3} \textit{Universit\'e d'\'Evry Val d'Essonne, Laboratoire de Math\'ematiques et Mod\'elisation d'\'Evry.}}}

{\let\thefootnote\relax\footnotetext{{{\it Acknowledgement:} We are grateful to Leif Andersen for comments and discussions. The research of St\'{e}phane Cr\'{e}pey benefited from the support of the Chair Markets in Transition under the aegis of Louis Bachelier laboratory, a joint initiative of \'Ecole polytechnique, Universit\'e d'\'Evry Val d'Essonne and F\'ed\'eration Bancaire Fran\c caise. }}}

\maketitle

\rm
\begin{abstract}

In the aftermath of the 2007 global  financial crisis, banks started reflecting into derivative pricing the cost of capital and collateral funding through XVA metrics. Here XVA is a catch-all acronym whereby X is replaced by a letter such as C for credit, D for debt, F for funding, K for capital and so on, and VA stands for valuation adjustment.

This behaviour is at odds with economies where markets for contingent claims are complete, whereby 
trades clear at fair valuations and 
the costs for capital and collateral are both irrelevant to investment decisions. 

In this paper, we set forth a mathematical formalism for derivative portfolio management in incomplete markets for banks. 
A particular emphasis is given to the problem of finding optimal strategies for retained earnings which ensure a sustainable dividend policy.
\end{abstract}

\def\keywordname{{\bfseries Keywords:}}
\def\keywords#1{\par\addvspace\baselineskip\noindent\keywordname\enspace
\ignorespaces#1}\begin{keywords}
Market incompleteness, counterparty risk, cost of capital, cost of funding, variation margin, initial margin, FVA, MVA, KVA.
\end{keywords}

\newpage
\tableofcontents
\newpage 

\section{Introduction}

In the aftermath of the financial crisis, regulators launched in a major banking reform effort aimed at securing the financial system by raising collateralisation and capital requirements and by pushing toward centrally cleared trading in order to mitigate counterparty risk. The resulting Basel III Accord as been enshrined in law and is being followed by a stream of other regulatory reforms such as the Fundamental Review of the Trading Book (FRTB, see \citeN{BISFRTB}).

The quantification of market incompleteness based on XVA metrics is emerging as the unintended consequence of the banking reform. Here XVA is a catch-all acronym whereby X is replaced by a letter such as C for credit, D for debt, F for funding, K for capital and so on, and VA stands for valuation adjustment. Interestingly enough, the new regulatory framework has been designed on the premise that the financial system can be secured by elevating capitalisation requirements for banks and forcing them to collateralise trades as much as needed to virtually eliminate counterparty credit risk, based on the assumption that costs of capital and funding for collateral are both irrelevant to investment decisions and do not give rise to material systemic inefficiencies. However, this conclusion is only valid under the complete market hypothesis. The impact of market incompleteness is exacerbated by the new regulatory environment, motivating investment banks to rewire trading strategies around the optimisation of XVA metrics, which precisely quantify the impact of market incompleteness. 
 
\subsection{Market Incompletenesses}\label{ss:mi}

The hypothesis of market completeness in \citeN{AD1954}, followed by the work of \citeN{BS73} and \citeN{Merton73}, represents a milestone for the development of markets for contingent claims as it indicates that derivative pricing can be based on replication arguments. This leads to remarkable simplifications: if all trades are perfectly replicable, then each trade is valued at its price of replication, independent of the endowment and of any other entity specific information. In complete markets, economic agents transact symmetrically and there is no distinction between price-maker and price-taker. There is also no distinction between 
fair valuation and entry price of a derivative contract. 
The most glaring omission in a complete market model for a bank is a justification for shareholders' capital at risk as a loss absorbing buffer. Derivative valuation adjustments such as cost of capital and cost of funding would be rigorously zero under the assumption of complete markets, as follows from the \citeN{MM1958} theorem. 

Since the crisis, market participants came to the realisation that costs of incompleteness are material and started to reflect these effects into entry prices. 
The primary focus was on hedging strategies for counterparty risk which are typically imperfect. Counterparty risk is the financial risk arising as a consequence of client or own default. The risk of financial loss as a consequence of client default is hard to replicate since single name CDS instruments are illiquid and are typically written on bonds, not on swaps with rapidly varying value. Own default of the bank is even harder (not to say impossible) to hedge since, 
to hedge it, a bank would need to be able to freely trade its own debt. Even so,
the shareholders of the bank could not effectively monetise the {hedging benefit}, which would be hampered by bankruptcy costs. In this situation, the assumptions of the Modigliani-Miller theorem do not hold and shareholders' decisions in general depend on the {funding strategy} of the bank.

In the case of counterparty credit risk, incompleteness is a complex phenomenon as it has capital structure implications and is perceived differently by shareholders and creditors. Counterparty credit risk is related to cash-flows or valuations linked to either counterparty default or the default of the bank itself. Shareholders are prevented by laws based on principles such as pari-passu from executing certain classes of trades that would transfer wealth away from creditors. Moreover, wealth transfers from shareholders to creditors cannot be offset in full by managers as banks are intrinsically leveraged and one cannot synthetically have shareholders become bond-holders and transform a bank into a pure equity entity. Interestingly enough, the latter is a trade on which the classical proof of the Modigliani-Miller theorem is based. Hence, only weaker statements will hold in the case of banks.
  
The discounted expectation of losses due to counterparty or own default are respectively known as CVA (credit valuation adjustment) and DVA (debt valuation adjustment). Regulators privilege collateralisation as the form of counterparty risk mitigant. However, the credit risk of the bank itself induces additional costs to fund margins that are required to mitigate counterparty risk:  variation margin tracking the mark-to-market of the portfolios of the bank with its counterparties and initial margins of the bank set as a cushion against gap risk, i.e.~the possible slippage between the portfolio and the variation margin of a defaulted party during the few days that separate the default from the liquidation of the portfolio. The discounted expectations of the cost of funding cash collateral which can be rehypothecated is known as funding valuation adjustment (FVA), while the cost of funding segregated collateral to be posted as initial margin is the margin valuation adjustment (MVA).
 
The CVA, FVA and MVA flow into a reserve capital account (RC), which is held against expected counterparty default and funding losses. As losses are realized, reserve capital may deviate from its theoretical equilibrium level given by the expectation of future losses, a value we call target reserve capital (TRC). Banks are required to hold economic capital (EC) sufficient to absorb exceptional losses. Economic capital is typically computed to be the expected shortfall with confidence level 97.5\% of the one-year-ahead loss. Since economic capital needs to be remunerated at the hurdle rate, bank clients are asked to pay an additional capital valuation adjustment (KVA), which then flows into the dividend stream. The KVA is not treated as reserve capital, but rather as a retained earning which contributes to economic capital.


\subsection{Contents of the Paper}

Most of the mathematical finance incomplete markets literature can be viewed as providing ways of selecting the ``right'' risk-neutral probability measure, under which financial derivatives are fairly valued as conditional expectation of their future cash-flows discounted at the risk-free rate. This includes utility indifference pricing, risk-minimisation and minimal martingale measures, utility maximisation and minimisation over martingale measures, good-deal pricing, market-consistent valuation, probability distortions, etc. (see e.g. \citeN{HodgesNeuberger89}, \citeN{Schweizer01}, \citeN{Rogers01dualityin}, \citeN{CochraneSaaRequejo00}, \citeN{Madan2014}). 


In this paper, we develop a mathematically rigorous, economically sustainable and numerically tractable XVA methodology, in two steps. First, a risk-neutral XVA (the target reserve capital TRC) is computed under a given a risk-neutral measure, possibly selected by one of above-mentioned procedures or simply by a pragmatic calibration approach delegating measure selection to market fit. Second, a risk-adjustment is computed on top of the TRC in the form of the KVA.

Part \ref{s:setup} provides an intuitive balance-sheet presentation of the main concepts  and ideas of the paper. 

Interestingly enough, Basel III/IFRS 9 do not contain a principled approach to cost of capital.
Adapting to banks the insurance principles of Solvency II and IFRS 4 Phase II, Part \ref{s:coc} introduces a precise and economically well grounded definition of the KVA as the cost of remunerating shareholders capital at a sustainable hurdle rate throughout the whole life of the portfolio (or until the default of the bank happens). 

The companion task is a proper modeling of the derivative portfolio loss process required as input in the economic capital and KVA computations. For this, a dynamic modeling of reserve capital and target reserve capital are required. This is done in Part \ref{p:rc} based on risk-neutral valuation principles, under the assumption
that trading strategies are self-financing from the viewpoint of shareholders of the bank (but may have the side-effect of transferring wealth between clients and bank creditors).
As compared with insurance, one difficulty specific to banks is that, in the case of derivative portfolios, capital is a form of funding which is intertwined and to a large extent fungible with debt. While insurance portfolios have only a KVA-like metric called risk margin (RM), banks have several other metrics such as FVA and MVA that are related to funding the collateral involved in OTC derivative transactions.
The intertwining between the FVA and EC metrics leads to a forward-backward stochastic differential equation (FBSDE) for the reserve capital and target reserve capital.
By doing so, we also improve the FVA models in \citeN{AAI2015} or \citeN{CrepeySong15} (see also \citeN{AlbaneseAndersenLongPaper2015} and \citeN{AndersenDuffieSong2016}).

Part \ref{s:impl} implements this XVA methodology by means of nested Monte Carlo simulations that are used for solving
the (RC,TRC) FBSDE iteratively. The ensuing loss process  $L=\Theta-\cV$ is then plugged as input data of the KVA computations. The paper is concluded by two case studies.

A list of acronyms is provided after the bibliography.


An announcement of the present work is in the short article by \citeN{AlbaneseCaenazzoCrepey15risk}. 


\part{Balance-Sheet Analysis}\label{s:setup}

\begin{figure}[h!]
\begin{center}
\begin{picture}(0,0)%
\includegraphics{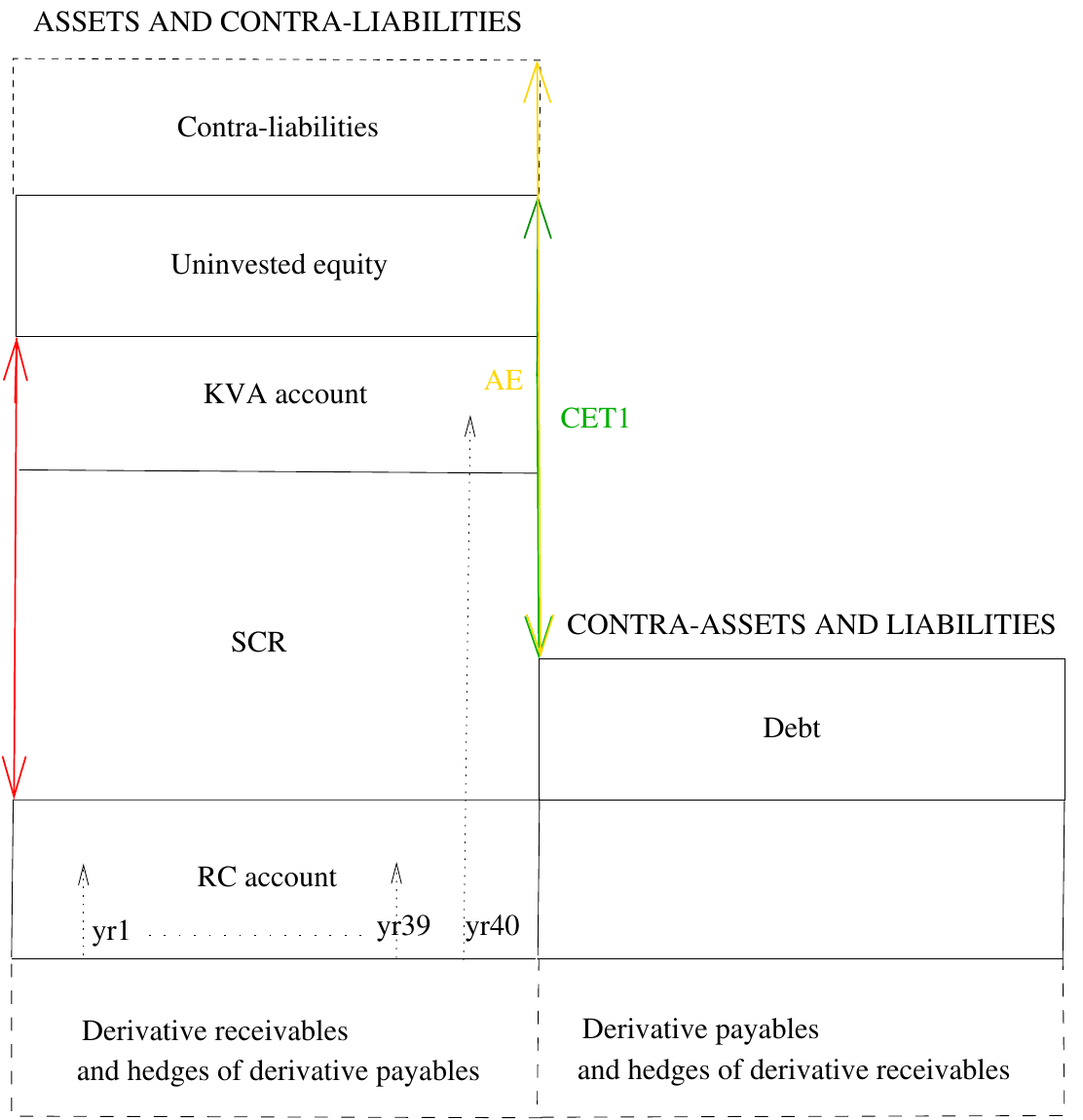}%
\end{picture}%
%
%
\setlength{\unitlength}{2171sp}%
\begingroup\makeatletter\ifx\SetFigFont\undefined%
\gdef\SetFigFont#1#2#3#4#5{%
  \reset@font\fontsize{#1}{#2pt}%
  \fontfamily{#3}\fontseries{#4}\fontshape{#5}%
  \selectfont}%
\fi\endgroup%
\begin{picture}(9695,10021)(153,-10408)
\put(5941,-8401){\makebox(0,0)[lb]{\smash{{\SetFigFont{9}{10.8}{\rmdefault}{\mddefault}{\updefault}{\color[rgb]{0,0,1}{\bf TRC} $= $UCVA+FVA+MVA}%
}}}}
\put(436,-5341){\makebox(0,0)[lb]{\smash{{\SetFigFont{11}{13.2}{\rmdefault}{\mddefault}{\updefault}{\color[rgb]{1,0,0}$\mbox{{\bf EC}}=\max\Big(  \mbox{ES}_{97.5\%} \big( \Delta_{1yr}(-\mbox{CET1})\big),\mbox{KVA}\Big)$}%
}}}}
\end{picture}%

\end{center}
\caption{Stylized balance sheet of an investment bank. 
Contra-liabilities at the left top are shown in a dashed box because
they should be ignored
from the point of view of the shareholders (they only contribute to AE, not to CET1). 
Financial payables (assets) and receivables (liabilities) at the bottom are shown in dashed boxes at the bottom because they
constantly match each other under our back-to-back market hedge assumption. The TRC (resp.~EC) computation is the goal of a reserve capital pricing (resp.~economic risk) model. The size of the KVA account is deduced from EC by the KVA formula and SRC is deduced from them as (EC$-$KVA). At each new deal, the RC account is refilled by corresponding incremental TRC amounts. Between deals, the RC account is gradually depleted by counterparty default risk and funding losses as they occur. 
The difference in size between the left and the right column in the figure corresponds to CET1. 
The dotted arrows represent RC losses in ``normal years 1 to 39'', 
and in an ``exceptional  year 40'' with full depletion of the RC, SRC and KVA accounts
(the numberings yr1 to yr40 are fictitious yearly scenarios in line with the 97.5\% confidence level used for the expected shorfall that underlies economic capital).  Whenever reserves are depleted and below the theoretical level, banks raise equity capital to realign them.}
\label{fig:bs}
\end{figure}
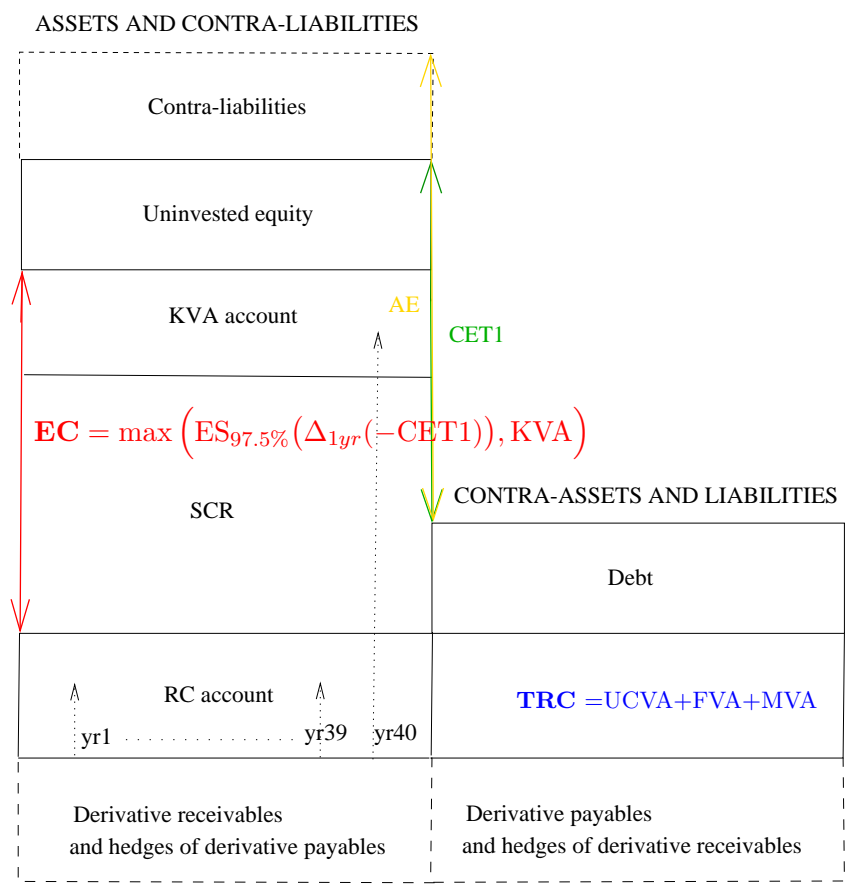

Figure \ref{fig:bs} represents a stylized balance sheet of an investment bank.  
The assets of a bank consist of reserve capital (RC), gross shareholder capital (capital at risk SCR and uninvested equity), risk margins (KVA), derivative receivables and hedges of derivative payables. Its liabilities are debt, derivative payables and hedges of derivative receivables. Contra-assets and contra-liabilities are respective assets and liabilities deductions, which will play a key role in this paper.

A defaultable entity is characterised by leverage and has at least two different classes of stakeholders: shareholders and creditors. Shareholders have the control of the firm and are solely responsible for investment decisions up until the time of default. At default time, shareholders are wiped out. Creditors instead have no decision power until the time of default, but are protected by laws such as pari-passu forbidding certain trades that would trigger wealth transfers from them to shareholders. 
Hence, there are two distinct but intertwined sources of market incompleteness: 
\begin{itemize}
\item Counterparty risk and market risk cannot be perfectly replicated,
\item Managers cannot offset wealth transfers from shareholders to bondholders and shareholders cannot realistically acquire all bank debt.
\end{itemize}
%
In order to focus on counterparty risk and XVAs, we assume throughout the paper that the market risk of the bank is perfectly hedged by means of back-to-back trades.
Under this assumption,
derivative receivables and hedges of derivative payables are always a perfect match to derivative payables and hedges of derivative receivables. In addition,
in case a new deal is traded, the gross shareholder capital of the bank and its debt remain constant. Hence, in case a new deal is traded,
\beql{e:deltaal}
\Delta(\mbox{Assets}-\mbox{Liabilities})=\Delta { \rm RC}+ \Delta { \rm KVA}.
\eeql

\section{Fair Valuation}\label{ss:bs}

A stylized balance sheet equation of a bank can be stated as (see Figure \ref{fig:bs}):
\beql{e:bsheet}
&\mbox{Assets}-\underbrace{(\mbox{UCVA}+\mbox{FVA}+\mbox{MVA})}_{\mbox{Contra-assets}}
=\\&\qqq\mbox{Liabilities}-
\underbrace{(\mbox{FTDDVA}+{{\rm CVA}_{\rm CL}}+\mbox{FDA}+\mbox{MDA})}_{\mbox{Contra-liabilities}}  + { \rm AE},
\eeql 
where:
\begin{itemize}
\item Assets and contra-assets, liabilities and contra-liabilities, {FVA} (funding variation adjustment) and {MVA} (margin variation adjustment)  have been introduced before;
\item  UCVA is the unilateral CVA pricing the cash-flows valued by the CVA over an infinite time horizon; 
\item FTDDVA is the first-to-default DVA pricing the cash-flows valued by the DVA until the first default time of the bank and each considered counterparty, and we define similarly the FTDCVA, i.e.~first-to-default CVA;
\item  ${\rm CVA}_{\rm CL}$ is the difference
$(\mbox{UCVA}-\mbox{FTDCVA});
$\item 
 FDA (akin to the DVA2 in \citeN{HullWhite13d}) 
and MDA are further contra-liabilities respectively equal to the FVA and the MVA (see {\shortciteN{AlbaneseAndersenLongPaper2015}} and 
\shortciteN{AAI2015});
\item  AE is the accounting equity of the bank.
\end{itemize} 

In virtue of the definition of ${\rm CVA}_{\rm CL}$ and of the cancellations between the funding terms, the equation \qr{e:bsheet} can be simplified as follows:
\beql{e:bsheet-exit}
\mbox{Assets}- \mbox{FTDCVA} 
=\mbox{Liabilities}-
  \mbox{FTDDVA}  + { \rm AE}.
\eeql

In case a new deal is traded, taking the difference between the balance-sheet equation \qr{e:bsheet-exit} at deal time and right before the deal yields, using \qr{e:deltaal}:
\bel
\Delta { \rm AE}=\Delta { \rm RC}+ \Delta { \rm KVA}-(\Delta\mbox{FTDCVA} - \Delta\mbox{FTDDVA}) .
\eel
If there is no risk or no capital at risk or if capital at risk is not remunerated, i.e.~in the absence of KVA,
then the  incremental reserve capital
needed from clients 
in order to keep accounting AE stable is 
\beql{e:bsheet-exit-incr}
\Delta { \rm RC}= \Delta\mbox{FTDCVA} - \Delta\mbox{FTDDVA}.
\eeql
Note that this formula involves the first-to-default CVA and DVA, where the counterparty default losses are only considered until the first occurrence of a default of the bank or its counterparty in the deal. This is consistent with the fact that later cash flows will, as first emphasised in \citeN{DuffieHuang}, \citeN{BieleckiRutkowski2002} and \citeN{BrigoCapponi2010}, not be paid in principle. The formula is symmetrical in that it also corresponds to the negative of the analogous quantity considered from the point of view of the counterparty of the bank.
In this sense, the symmetrical adjustment \qr{e:bsheet-exit-incr} to the mark-to-market of a deal corresponds to
fair valuation.

\section{Entry Price}\label{ss:ep}

However, such a fair value of the deal cannot correspond to its entry price unless one is in a complete market where contra-liability cash-flows can be hedged and therefore monetized by the shareholders, and where counterparty (and market) risk can be hedged as well so that no capital needs be put at risk by the shareholders and remunerated to them at a hurdle rate.

In fact, contra-liability cash-flows
only benefit to the creditors of the bank, in the form of an increased recovery rate. As a consequence,
they should be ignored in entry prices, which must be aligned to the interests of the shareholders. 

Following up to these considerations, regulators have de-recognised the DVA as a contributor to Core Equity Tier I capital (CET1), 
the metric meant to represent the fair valuation of shareholder capital. 
Recently, the financial accounting standards board (FASB) has redefined 
the generally accepted accounting principles (GAAP)
and stated that the DVA should not contribute to reported earnings. Regulators went even further and decided that the CVA should be computed unilaterally to ensure that it is a monotonic function of the bank credit spread. As we shall will demonstrate
in \sr{ss:csaxvaeq} (cf. \qr{e:ucvatau} and \qr{e:cva-fvabis}), this decision can be modelled by postulating that shareholders are liable to pay UCVA$_\tau$ 
to bank bondholders
at the default time $\tau$ of the bank.

The corresponding contra-liabilities myopic, shareholders-centric balance sheet equation of the bank  appears as (cf.~the full-balance sheet equation \qr{e:bsheet}  and Figure \ref{fig:bs}):
\beql{e:bsheet-entry}
\mbox{Assets}-\underbrace{(\mbox{UCVA}+\mbox{FVA}+\mbox{MVA})}_{\mbox{Contra-assets}}
=\mbox{Liabilities} 
+  {\rm CET1}.
\eeql
In case a new deal is traded, 
if there was no KVA,
then, using \qr{e:deltaal}, this would result in
\bel
 \Delta {\rm CET1}=\Delta\mbox{RC}-( \Delta\mbox{UCVA} + \Delta\mbox{FVA}+\Delta\mbox{MVA}).
\eel 
Hence, the incremental amount of reserve capital 
that would keep stable the CET1 net of the KVA account is
\beql{e:bsheet-entry-incr-delta-prel}
\Delta\mbox{RC}= \Delta\mbox{UCVA} + \Delta\mbox{FVA}+\Delta\mbox{MVA}= \Delta\Theta ,
\eeql
where the target reserve capital (TRC) is defined as the sum of the contra-assets, i.e.~
\beql{e:bsheet-entry-incr-delta-post}
\Theta=\mbox{UCVA}+\mbox{FVA}+\mbox{MVA}.
\eeql

However, shareholders want their risk to be remunerated at a hurdle rate $h,$ which constitutes an additional cost passed to the clients, measured by a suitable KVA metric. In the end the add-on
to the entry price for a new deal, called funds transfer pricing (FTP),
 is
\beql{e:bsheet-entry-incr-delta}
{\rm FTP}= \Delta\Theta +\Delta\mbox{KVA},
\eeql
where $\Delta\mbox{KVA}$ is the incremental risk margin of the new deal.
 
As we shall understand in more detail later, the TRC
can be interpreted as the risk-neutral
contribution to the FTP \qr{e:bsheet-entry-incr-delta},
accordingly we sometimes call it risk-neutral XVA.
The KVA can be interpreted as a risk-adjustment.

Note that banks are market makers and, as such, they are price makers. Bank clients are price takers willing to accept a loss in a trade for the sake of receiving benefits which become apparent only once one includes their real investment portfolio, which cannot be done explicitly in a pricing model. 
In an asymmetric setup with a price maker and a price taker, the price maker  passes his costs to the price taker.
Things get a bit tricky for bilateral trades between two financial intermediaries.
In this case each party will try to have the other pay its costs. This in general would result in no deal but in practice translates into a shared loss somewhere in the middle of the range, depending on which party has the strongest contractual power.

\section{Dynamic Setup}\label{ss:msetup}

The above balance sheet reasoning depicts a framework which is qualitatively correct, but is insufficient to provide a model of CET1 fluctuations needed as an input into KVA calculations, or to quantify  finer effects such as the intertwining between economic capital and FVA that results from the possibility to use economic capital as a source of funding. For this, a classical continuous-time mathematical finance implementation of the above ideas is required, which is the purpose of the remainder of the paper.

Let $(\Omega,\cA,\gg,\mathbb{Q}),$ with $\gg=(\G_t)_{t\in\R _+}$, represent a suitable
stochastic basis, with expectation (resp.~conditional expectation) denoted by $\mathbb{E}$ (resp.~$\mathbb{E}_t$), such that all the processes (resp.~random times) of interest are $\gg$ adapted (resp.~$\gg$ stopping times). 

Target reserve capital valuation will be based on a no-arbitrage martingale condition with respect to a risk-neutral pricing measure. By contrast,
economic capital and KVA  assess risk and its cost,
which should be computed under the historical probability measure  (at least in principle, let aside implementation issues). 
Hence, the probability measure $\mathbb{Q}$ of choice in Part \ref{s:coc} 
typically differs from the one to be used in Part \ref{p:rc}.



The connection between the above balance-sheet analysis and 
the dynamic analysis that follows sits in
the way this dynamic model should be used in practice. Namely, time 0 in the dynamic model represents the time a new deal is considered. For the purpose of computing incremental XVAs for this new trade, the (back-to-back hedged) derivative portfolio of the bank, including the new trade, is modelled on a run-off basis
until its final maturity $T$ or the bank's default time $\tau$.
Hence, the time horizon of the model is $\db=\tau\wedge T.$

We denote by $\loss=\Theta-\cV$ 
the 
loss process (cumulative realized 
loss and profit) of the back-to-back hedged portfolio. 
An accrued loss $y$ represents the value $\widetilde{\Theta}_0-\widetilde{\cV}_0$ attained
at time 0 by the loss process of the portfolio without the new deal. Hence, an initial condition $\loss_0=y$ states that
$$\cV_0=\Theta_0-(\widetilde{\Theta}_0-\widetilde{\cV}_0)
=\widetilde{\cV}_0+( \Theta_0 -\widetilde{\Theta}_0).$$
In other words, the incremental amount $\Delta {\rm  RC} =\cV_0-\widetilde{\cV}_0$ of reserve capital that should be charged to the client for the new deal at time 0 equals $\Theta_0-\widetilde{\Theta}_0=\Delta {\rm  TRC},$ 
which is consistent with the balance sheet formula \qr{e:bsheet-entry-incr-delta-prel}. 

Similarly, in order to adjust the size of the KVA account from the starting value $\widetilde{{\rm KVA}}_0$ to the new value ${{\rm KVA}}_0$ reflecting the inclusion of the new deal, the bank must 
pass on to the client an incremental KVA amount equal to the difference $\Delta {\rm KVA}={\rm KVA}_0-\widetilde{{\rm KVA}}_0.$
In the end, the total XVA charge passed on to the client of the new deal is given by
the FTP formula (\ref{e:bsheet-entry-incr-delta}).

\part{Cost of Capital}\label{s:coc}

The cost of capital is the cost of remunerating the bank shareholders in order to compensate the latter for placing capital at risk. 

If markets were complete and all payoffs were perfectly replicable, then no capital would be at risk and no compensation would be required. However, in incomplete markets this is not the case. Shareholders' capital is exposed to financial risk and the risk compensation needs to be sourced from clients in proportion to a measure for capital consumption.

The economic capital (EC) of a bank,  which is its loss-absorbing resource devoted to cope with exceptional losses beyond reserve capital, consists of the sum between shareholders' capital at risk (SCR) and retained earnings (\mbox{KVA}). Regarding economic capital,
the FRTB talks about a VaR replacement, defining EC as expected shortfall (ES) with 97.5\% confidence
(cf.~\citeN{HEAD}). This however refers to market risk, not CET1.
On the other hand, Basel II Pillar II defines economic capital as the year-on-year 99\% VaR of ($-$CET1).
Putting together the two, we arrived to our updated variant of EC defined as 
\beql{e:yyeces}\mbox{The year-on-year } 97.5\% \mbox{ ES for } (-\mbox{CET1})\eeql
(see \qr{e:ec} for a refined technical definition also integrating a Solvency-like accounting condition).
The interpretation of this expected shortfall   
is that EC is seized to a level such to absorb average losses deriving by extreme levels of depletion of reserve capital occurring in the worst out of every forty years (cf.~Figure \ref{fig:bs}).

The level of compensation required by shareholders on SCR is driven by market considerations. Typically, investors in banks expect a hurdle rate $h$ of about $10\%-12\%$.  
When a bank charges cost of capital to clients, these revenues are accounted for as profits. Unfortunately, since prevailing accounting standards for derivative securities are based on the theoretical assumption of market completeness, they do not envision a mechanism to retain these earnings for the purpose of remunerating capital across the entire life of transactions, that can be as long as decades. In complete markets, there is no justification for risk capital. Hence, profits are immediately distributable.  A strategy of earning retention beyond the end of the ongoing accounting year (or quarter) is still possible as in all firms, but this would be regarded as purely a business decision, not subject to financial regulation under the Basel III Accord.

An argument as of why banking retained earnings policies should be regulated is related to the so called ``leverage ratchet effect'' and is discussed in \citeN{ADMHP2013}. According to their paper, shareholders have a tendency to ratchet up leverage even if this comes to detriment to the value for the firm as a whole. In the case of banks, the ratchet effect is compounded by the mechanics of capital remuneration under current accounting rules:
since earnings received from clients to remunerate capital are accounted for as day-one profits, they are immediately distributable unless so decided by the bank board of directors. This leads to an explosive instability characteristic of a Ponzi scheme. Initially, derivative markets grow driven by trades used for hedging purposes and essential for the efficient functioning of the economy. As volumes for derivative markets keep growing, the increasing need of financing capital attracted by these derivative portfolios leads to the onset of a phase where new derivative asset classes are created for speculation purposes only.
 
For instance, if a bank starts off today by entering a 30-year swap with a client, the bank books a profit. Assuming the trade is perfectly hedged, the profit is distributable at once. The following year, the bank still needs capital to absorb the risk of the 29-year swap in the portfolio. But how can the bank remunerate shareholders if the profits from this trade have already been distributed the previous year? Simple: {lever up! The bank sells and hedges another swap, books a new profit and distributes the dividend to shareholders that are now posting capital for both swaps. As long as trading volumes grow exponentially, the scheme self-sustains. When exponential growth stops, the bank's return on equity crashes.

The financial crisis of 2007-2008 can be largely explained along these lines (see Figure \ref{fig:ponzi}). In the aftermath of the crisis, the first casualty was the return on equity for the fixed income business as profits had already been distributed and market-level hurdle rates could not be sustained by portfolio growth. 
\begin{figure}[h!]
\centering
\includegraphics[width=0.49\textwidth,height=0.27\textheight]{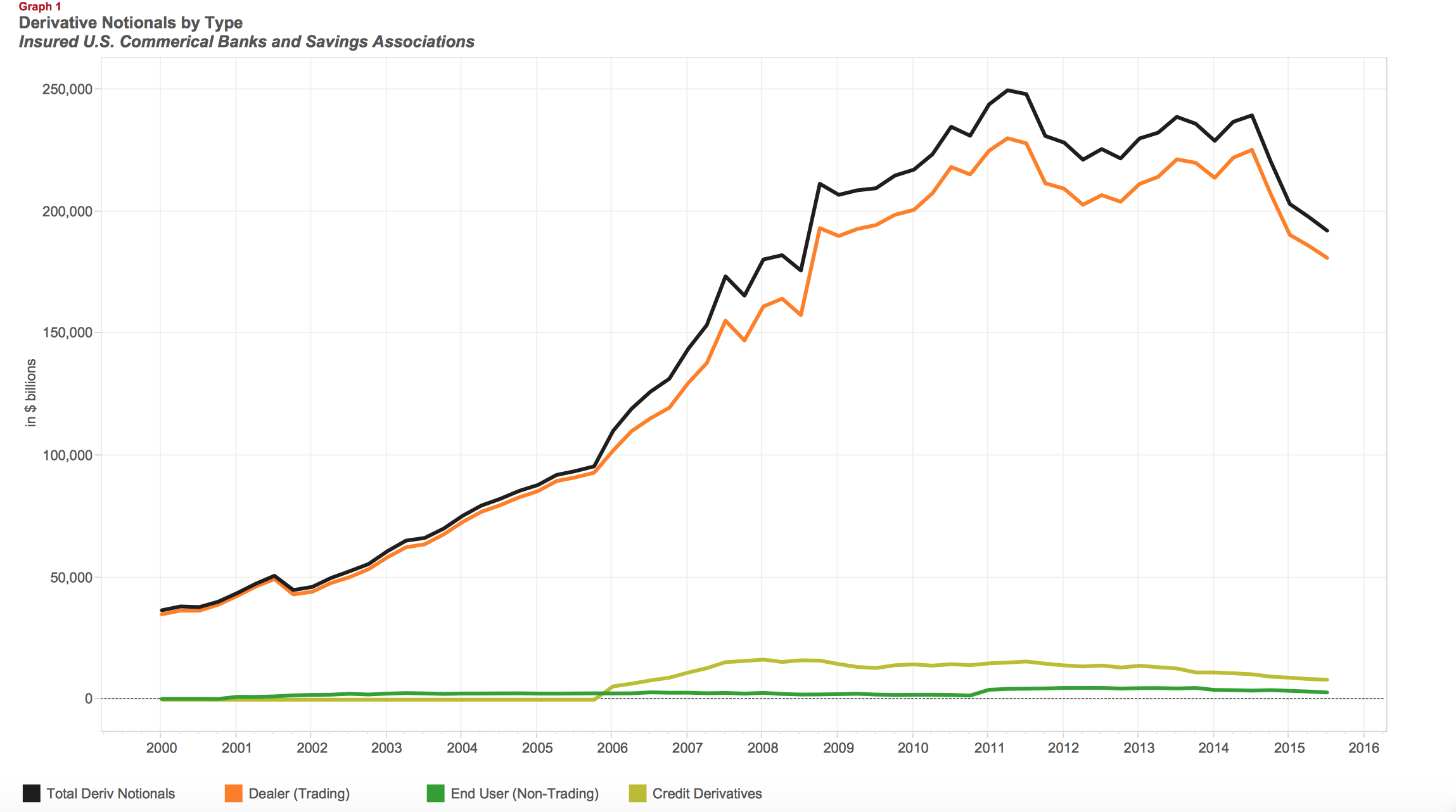}
\caption{Ponzi scheme in the last financial crisis
(source: OCC Q3 2015 Quarterly Bank Trading Revenue Report).}
\label{fig:ponzi}
\end{figure}
Interestingly enough however, in the insurance domain, \citeN{Solvency2}, unlike Basel III, does regulate the distribution of retained earnings through a mechanism tied to so called ``risk margins'', which directly inspires the KVA metric discussed in this article. 
Also, the accounting standards set out in 
IFRS 4 Phase II
(see \citeANP{IFRS13} (\citeyearNP{IFRS13}, \citeyearNP{IFRS4Phase2ED})
are consistent with Solvency II and include a treatment for risk margins that has no analogue in the banking domain. See \citeN{WuthrichMerz13}, \citeN{EiseleArtzner11} and \citeN{SalzmannWuthrich2010} regarding the risk margin and cost of capital actuarial literature.

The purpose of this part is to discuss a framework for assessing cost of capital (KVA) for a bank, pass it on to the bank's clients and distribute it
gradually to the bank's shareholders through a dividend policy which would be sustainable even in the limit case of a portfolio held on a run-off basis, with no new trades ever entered in the future. Solvency II requires in addition that EC must be higher than the cost of capital (i.e.~the KVA). In this paper, we compute the KVA also respecting this additional Solvency II constraint.

\section{KVA Equations}\label{ss:cocequ}

Let $\RM =\RM_t (\loss)$ denotes the 97.5\% {conditional expected shortfall} 
of the one-year-ahead \lp ($\loss _{t+1}-\loss _{t}$) of the bank 
computed on a going-concern basis, i.e.~conditionally to bank survival,
that is, writing $\Q_t$ for $\E_t\ind$:
\beql{e:rm}
\RM_t(L) = \frac{\E_t  \big[ ( \loss _{t+1}-\loss _{t} )\ind_{\{\loss _{t+1}-\loss _{t} \ge \VaR_t(\loss) \}\cap\{\tau>t+1\}}\big] }{\Q_t  \big[ \{\loss _{t+1}-\loss _{t} \ge \VaR_t(\loss) \}  \cap\{\tau>t+1\}  \big]},
\label{eq:EC}
\eeql
where
\beql{e:var}
&\VaR_t(\loss)= \inf \Big\{ \ell : \frac{\Q_t \big[ \{ \loss _{t+1}-\loss _{t}   \geq  \ell \} \cap \{\tau>t+1\} \big]}{\Q_t (\{\tau>t+1\})} \leq 2.5\% \big\}.
\eeql

For KVA computations entailing capital projections over decades, an equilibrium view based on Pillar II  
economic capital (\mbox{EC}) is more attractive than the ever-changing Pillar I regulatory charges supposed to approximate it, see \citeN{Pykhtin12}. However, Pillar I regulatory capital requirements can be incorporated into our approach if wished by replacing
$\RM$ by its maximum with the regulatory capital pertaining to the portfolio.

We denote by $r$ a $\gg$ progressive overnight index swap (OIS) rate  process and by $\beta_t=e^{-\int_0^t r_s ds}$ the corresponding discount factor. The OIS rate is altogether the best market proxy for a risk-free rate and the reference rate for the remuneration of cash collateral. Let $C\geq \mbox{ES}$ represent a putative economic capital process for the bank.
To compute the KVA we ask the following question: what should be the process for the size $K=K_t(C)$ of a KVA account, accruing at the OIS rate $r$, 
in order to generate an average (expected) remuneration to shareholders equal to $h(C_t - K_t) dt$ at each point in time {$t\in[0,\tb]$}?
The reason why $(C_t - K_t)$ rather than $C_t$ appears in this formula is that retained earnings are loss-absorbing and therefore part of the economic capital. Therefore, shareholders' capital at risk, which by assumption is remunerated at the hurdle rate $h,$ only corresponds to the difference $(C_t - K_t).$

Summarizing, we are looking for a process $K_t$
that satisfies the terminal condition $K_{\tb} =0$ and obeys the equation
\beql{e:truek}
 dK _t+\big(h ( C_t-K _t)-r K_t \big)dt \mbox{ is a local martingale on } [0,\tb]  .\eeql
The differential specification \qr{e:truek} is in the form of a linear backward stochastic differential equation (BSDE)\footnote{We refer the reader to \citeN{Crepey12} for an introductory financial modeling BSDE reference.}, equivalent to the integrated form \qr{e:k} below. 
As we demonstrate in Lemma \ref{l:wp} below, the solution $K=K_t(C)$ to this equation is unique. Furthermore, since the equation is linear, its solution is given by the explicit KVA formula in \qr{e:kexpl}. 

But the difference $(C_t-K_t)$ represents shareholder capital at risk and must therefore be non-negative in order to satisfy a Solvency-like accounting condition (cf.~\sr{ss:cocopt}).
If one accounts for the corresponding constraint $C\geq  K ,$ then the BSDE becomes of non-linear type \qr{e:kva} with a Lipschitz coefficient.

We denote by
 $\cH^p$ the space of ${\cdot}^p$-integrable processes over $[0,\tb],$ for any $p\geq 1.$

\bl\label{l:wp} Consider the following BSDEs:
\beqa
&&\mbox{KVA}_t=  \E_t \int_t^\tb \big( h \max ( \RM _s,  {\mbox{KVA}}_s ) -  (r_s  +h){\mbox{KVA}}_s\big) ds\sp t\in [0,\tb] \label{e:kva}\\
&&K_t=  \E_t \int_t^\tb \Big(h  C_s-(r_s  +h)K_s \Big)ds\sp t\in [0,\tb] ,\label{e:k}
\eeqa
to be solved for respective processes $K$ and $\mbox{KVA}.$
Assuming that $r$ is bounded from below  
and that $\RM $ (respectively $C$) and $r$ are in $\cH^2$, then the BSDE \eqref{e:kva}  (respectively \eqref{e:k}) is
well posed in $\cH^2,$ where well-posedness includes existence, uniqueness,
comparison and the standard a priori bound and error BSDE  estimates. The $\cH^2$ solution $K$
to \eqref{e:k} admits the explicit representation
\beql{e:kexpl}
K_t =h \E_t \int_t^\tb e^{-\int_t^s (r_u +h) du}  C_s ds ,\,t\in[0,\tb].
\eeql\el
\proof Let, for $k\in\R ,$
\beql{e:f}
f_t (k) = h \big(\max ( \RM _t ,  k)-k\big) -   r_t   k =h \max\big( \RM _t ,  k\big) -  (r_t   +h)k,
\eeql
which is the coefficient of the KVA BSDE \qr{e:kva}.
For any real $k,k'\in\R$ and $t\in[0,\tb],$ it holds
\bel&\big( f_t(k) - f_t(k')\big) (k-k')=
-(r_t +h)(k-k')^2 +h\big(\max ( \RM _t ,  k )-\max ( \RM _t ,  k')\big) (k-k'))\\&\qqq \leq - r_t  (k-k')^2 \leq   C  (k-k')^2  \eel
(having assumed $r$ bounded from below), so that the coefficient $f$ satisfies the so-called monotonicity condition.
Moreover, for $|k|\leq \bar{k}$, we have that
\bel
|f_{\cdot}(k)-f_{\cdot}(0)|
&\leq  h \max\big( |\RM|   , \bar{k}\big)+ |h+r   | \bar{k}+ h \RM^+   .
\eel
Hence, assuming that $\RM $ and $r$ are in $\cH^2 ,$
the following integrability conditions hold:
\beql{e:as}\sup_{|k|\leq \bar{k}}|f_{\cdot}(k)-f_{\cdot}(0)|\in {\mathcal{H}}^1 ,\mbox{ for any } \bar{k}>0, \;\;\;\;\;{\rm and}\;\;\;\;\; f_{\cdot}(0)\in {\mathcal{H}}^2 .\eeql
Therefore, by application of classical BSDE results (see e.g.~\citeN[Sect.~5]{KrusePopier14}), the BSDE \eqref{e:kva} is well-posed in $\cH^2,$ where well-posedness includes existence, uniqueness,
comparison and the a priori bound and error BSDE estimates.
Even simpler computations prove the statement regarding the linear BSDE \eqref{e:k}. Hence, \eqref{e:k} is well-posed in $\cH^2$. Moreover, \qr{e:kexpl} obviously solves \eqref{e:k}.~\finproof\\

Note that, in case some exceptional ``40-yearly'' loss event occurs, according to the total loss-absorbing capacity standard TLAC (see \citeN{TLAC}), debt is converted into equity to replenish CET1 (cf.~Figure \ref{fig:bs}). Within our proposed treatment for risk margins, this conversion would replenish both shareholder capital at risk and the KVA. This explains why the stopping times of exceptional losses do not appear in the above 
KVA 
metric.

\section{The Solvency II Constraint on the KVA}\label{ss:cocopt}

Under the insurance Solvency capital requirement, economic capital is the sum between shareholder capital at risk (SCR) and risk margins (the insurance analog of the KVA). 
Some actuarial literature dwells with the puzzle according to which the calculation of the risk margins depends on economic capital projections in the future while economic capital itself depends on the risk margins, an apparently circular dependency (see e.g.~\citeN[Sect.~4.4]{SalzmannWuthrich2010}, \citeN{Robert13} and \citeN{EiseleArtzner11}). 

This paper addresses the problem of circular dependency as follows: First, we compute the economic capital according to some risk measure, where the economic capital corresponds to the sum (\mbox{SCR} + \mbox{KVA}).
Then, we define  \mbox{KVA} using economic capital projections discounted at a hurdle rate  (cf.~\qr{e:kexpl}). The \mbox{SCR} is defined a posteriori as the difference $(\mbox{EC} - \mbox{KVA})$
(see Figure \ref{fig:bs}).
However, we need to account for the additional constraint that $\mbox{KVA}\leq \mbox{EC}$. Otherwise this would break the Solvency-like accounting condition 
$\mbox{SCR}\ge 0$ (discussed after \qr{e:adm})
and KVA would cease to be a supermartingale and to decrease as a function of the hurdle rate $h$. This additional constraint
is handled by considering the modified \mbox{KVA}
(i.e.~the \mbox{KVA}) defined in terms of the \mbox{KVA} BSDE \qr{e:kva}.

To emphasize the dependence of $K$ on $C$, we henceforth denote by $K(C)$ the solution \qr{e:kexpl} to the linear BSDE \qr{e:k}. In view of Lemma \ref{l:wp},
the process $K=K(C)$ defined by \qr{e:kexpl} is the unique sustainable dividend policy with hurdle rate $h$ associated with a putative economic capital process $C\in\cH^2$.
The set of admissible economic capital processes is \beql{e:adm}\cC=\{C\in 
\cH^2 ; C\geq \max( \RM  , K(C))\},\eeql
where $C\geq  \RM  $ is the risk acceptability condition and $C\geq K(C)  $  corresponds to a Solvency-like accounting condition, expressing a switching (or transferability) property of the bank to new shareholders if need be (cf.~the respective conditions (b) and (a) and their discussion in \citeN[pages 270 and 271]{WuthrichMerz13}
and see also the refined discussion in \sr{ss:solv}).
We define \beql{e:ec}\mbox{EC}=\max(\RM ,\mbox{KVA})\eeql
in $\cH^2,$ where KVA is the $\cH^2$ solution to the BSDE \qr{e:kva}.
\brem\label{r:ecr} In many cases as in our concluding figure \ref{fig:EC},
we have that $\mbox{EC} = \RM$. The inequality stops holding when the hurdle rate is high enough and the term structure of EC starts very low and has a sharp peak in a few years.
\erem
{Note that the KVA solution to \qr{e:k} also solves a linear BSDE \qr{e:k} (and is therefore given by the linear KVA formula \qr{e:kexpl}), but for the implicit data $C=\mbox{EC}$.}

The next result shows that $\mbox{EC}$ is a an optimal admissible economic capital process, optimal in the sense of smallest and with the cheapest ensuing cost of capital, which is shown to be nondecreasing in the hurdle rate $h.$

\bt Under the assumptions of Lemma \ref{l:wp}, we have:
\hfill\break {\rm \textbf{{(i)}}} $\mbox{KVA}=K(\mbox{EC}).$
\hfill\break{\rm \textbf{{(ii)}}} $\mbox{EC}=\min \cC, \mbox{KVA}=\min_{C\in\cC}K(C).$
\hfill\break{\rm \textbf{{(iii)}}} If $\mbox{ES}\geq 0,$ then $\mbox{KVA}$ is nondecreasing in $h.$
\et
\proof {\rm \textbf{{(i)}}}
$\mbox{KVA}$ is in $\cH^2$ and, by virtue of \qr{e:kva}, we have, for $t\in[0,\tb],$
\bel
&\mbox{KVA}_t=  \E_t \int_t^\tb\Big( h \max\big( \RM _s,  {\mbox{KVA}}_s\big) -  (r_s  +h){\mbox{KVA}}_s\Big) ds\\
&\qqq =  \E_t \int_t^\tb \Big(h  {\mbox{EC}}_s-(r_s  +h){\mbox{KVA}}_s \Big)ds  ,
\eel
so that the process $\mbox{KVA}$ solves the linear BSDE \qr{e:k} for $C=\mbox{EC}\in\cH^2$.
Hence $\mbox{KVA}=K(\mbox{EC})$ follows by uniqueness of an $\cH^2$ solution to the linear BSDE \qr{e:k} established in Lemma \ref{l:wp}.
\hfill\break{\rm \textbf{{(ii)}}} We just saw that
$ \mbox{KVA}=K(\mbox{EC}),$ hence
$$\mbox{EC}=\max(\RM ,\mbox{KVA})=\max\big(\RM ,K(\mbox{EC})\big).$$ 
Therefore
$\mbox{EC}\in \cC.$ Moreover, for any $C\in\cC$ (cf.~\qr{e:f})
\bel
& f_t(K_t(C))  = h \max\big( \RM _t,  K_t(C)\big) -  (r_t  +h)K_t(C) \leq h C_t - (h+r_t)K_t(C).
\eel
Hence, the coefficient $f$ of the KVA BSDE \qr{e:kva} never exceeds the coefficient of the linear BSDE \qr{e:k} when both coefficients are evaluated at the solution $K_t(C)$ of the linear BSDE \qr{e:k}. Since these are BSDEs with equal (null) terminal condition,
the comparison theorem applied to the
BSDEs \qr{e:kva} and \qr{e:k}
yields $\mbox{KVA}\leq K(C).$
Consequently, $\mbox{KVA}=\min_{C\in\cC} K(C)$
and, for any $C\in\cC,$
$$C\geq \max(\RM ,K(C))\geq \max(\RM ,\mbox{KVA})=\mbox{EC},$$
hence $\mbox{EC}=\min \cC$.
\hfill\break{\rm \textbf{{(iii)}}} If $\mbox{ES}\geq 0,$ then, as visible in \qr{e:f}, 
the coefficient $f_t(k)$ of the KVA BSDE \qr{e:kva} is nondecreasing in the hurdle rate parameter $h.$ So is therefore in turn the $\cH^2$ solution $\mbox{KVA}$ to \qr{e:kva}, by the comparison theorem applied to the BSDE \qr{e:kva} for different values of $h$.~\finproof

\part{Reserve Capital and Target Reserve Capital}\label{p:rc}

Having designed the XVA risk-adjustment methodology in the form of a KVA defined as cost of capital, the companion task is to specify a proper modeling  framework for the loss process $L$ to be used as input in these cost of capital computations. By doing so, we also improve the FVA models in \shortciteN{AAI2015} or \citeN{CrepeySong15} (see also \shortciteN{AlbaneseAndersenLongPaper2015} and \citeN{AndersenDuffieSong2016}).

In this part the probability measure $\mathbb{Q}$ denotes
a risk-neutral pricing measure calibrated to the market.

Liquidation and funding costs, which are primarily driven by counterparty risk, are hard to hedge in practice.
Reserve capital (RC) is used for dealing with expected liquidation and funding costs as they occur. Exceptional losses are accounted for by economic capital}.

At each trade, incremental {UCVA}, {FVA} and {MVA} amounts are charged to clients and flow into the RC account (cf.~\qr{e:bsheet-entry-incr-delta-prel}).

In between trades, liquidation and funding costs deplete the RC account, whereas the target reserve capital (TRC) 
fluctuates according to model valuation 
computed assuming the portfolio held under a run-off basis.

As a result, a discrepancy $L=\Theta-\cV$ develops between the reserve capital and its target value. Economic capital is required from the bank in order to deal with the fluctuations of this loss process $L.$
To make the connection with \qr{e:yyeces}, note that, 
in the case of a portfolio held on a run-off basis, gross equity, which is the sum between gross shareholder capital and the KVA, is constant.
For instance, if the risk increases on the market, i.e.~if
$\mbox{EC}=\mbox{SCR}+\mbox{KVA}$ increases, then this must be compensated by an equal decrease in uninvested equity. 
This should be contrasted with what happens at a new deal where, as mentioned before \qr{e:deltaal}, it is only the gross shareholder capital
(gross equity net of the KVA) that is constant, but an incremental KVA can be sourced from the client. In both cases, the debt of the bank is also constant.
 Hence,
under our run-off and back-to-back market hedge assumptions on the bank portfolio, the fluctuations of the process $L=\mbox{TRC}-\mbox{RC}$ are nothing but the fluctuations of ($-\mbox{CET1}$) (cf.~the CET1 defining equation \qr{e:bsheet-entry} and Figure \ref{fig:bs}).   

Now, for a proper modeling of the fluctuations of the loss process $L=\Theta-\cV$  (or, equivalently, of ($-\mbox{CET1}$)) required as input to our KVA computations, dynamic modeling of the reserve capital and of the target reserve capital are required.
%
%
Our modelling principle in this regard is 
 that, in order to prevent arbitrage, the 
loss process $L=\mbox{TRC}-\mbox{RC}$ should follow a risk-neutral local martingale.  

The principles of no arbitrage and risk neutral valuation have seen successive developments in \citeANP{BDF1931} (\citeyearNP{BDF1931}, \citeyearNP{BDF1937}), \citeN{MM1958}, \citeN{HarrisonPliska} and \citeN{Delbaen2005} and are surveyed in \citeN{Duffie}.  
As recalled in Part \ref{s:setup}, contra-liability entries (such as the DVA) price cashflows which occur at the time of bank default and have an impact on the worth of bank debt by affecting the recovery rate. However, only the interest of shareholders matters in bank's managerial decisions. Hence, in our setup, we postulate a risk-neutral local martingale for $L=\mbox{TRC}-\mbox{RC}$ when all the contra-liabilities are ignored.

First, we write a shareholders-centric self-financing condition ignoring contra-liability cash-flows, which results in a forward SDE for RC reflecting the counterparty default and margining funding losses. The initial condition for this SDE is
$
\cV_0=\Theta_0-y,$
where
the accrued loss $\loss_0=y$ is the negative of the initial endowment of the portfolio. 

As illustrated in a concrete setup in \sr{ss:csaxvaeq}, a no arbitrage risk-neutral local martingale condition on the \lp process $\loss=\Theta-\cV$ is then equivalent to a backward SDE for the $\Theta$ process with terminal condition $\Theta_\tb=0$, coupled with the forward SDE for $\cV$.

This leads us to a forward-backward SDE (FBSDE) to be satisfied by the pair of processes $(\cV,\Theta)$.
Moreover,
since economic capital can be used for funding purposes, there will be a feedback loop from economic capital into the FVA component of the TRC: 
The higher economic capital, the lower the FVA.

%
 

\section{Margining, Partial Hedging, Funding and Losses Realization Schemes} \label{s:pomapa}

The data of the $(\cV,\Theta)$ FBSDE reflect derivative portfolio margining, partial hedging, funding and losses realization schemes, which
we specify in this section.

We consider $(n+1)$ financial entities trading together, 
the bank and $n$ counterparties such as sovereigns, corporate entities and possibly other banks or financial insitutions, indexed by $i=1,\ldots, n,$ with default times $\tau_i$ and survival indicators $J^i=\ind_{[0,\tau_i)}$.  The bank
is also default prone, with default time $\tau$ and survival indicator $J=\ind_{[0,\tau)}$. We suppose that all these default times admit a finite intensity. In particular, defaults occur at any given $\gg$ predictable time with zero probability (but we do not exclude simultaneous defaults).

To mitigate counterparty risk, the bank and its counterparties exchange variation and initial margin. Variation margin typically consists of cash that is re-hypothecable, meaning that received variation margin can be reused for funding purposes, and is remunerated at OIS by the receiving party. Initial margin typically consists of liquid assets deposited in a segregated account, such as government bonds, which naturally pay coupons or otherwise accrue in value. The poster of the collateral receives no compensation, except for the natural accrual or coupons of its collateral.

As happens in practice in the current regulatory environment, the back-to-back market hedge of the derivative portfolio of the bank is assumed to be with other financial institutions and attracts variation margin at zero threshold. 
On top of its back-to-back market hedge, the bank may setup a counterparty risk (partial) hedge $\eta,$ which is a predictable and locally bounded vector process of dynamic positions in some hedging assets. We denote by  $\mathcal{M}$ the vector-valued gain process of unit positions in these hedging assets, assumed a risk neutral vector local martingale.
As explained in the comments following
the assumption 4.4.1 in \citeN[page 96]{BieleckiBrigoCrepeyHerbertsson13}\footnote{Or \citeN[Part I, Assumption 4.1]{Crepey2012bc} in article version.},
this assumption on $\cM$ rules out arbitrage opportunities in the market spanned by hedging instruments\footnote{Provided one restricts attention to hedging strategies resulting in a wealth process bounded from below (see
\citeN[Corollary 3.1]{BieleckiCrepeyRutkowski11} for a formal statement).}.
The counterparty risk hedge $\eta$ is a replicating strategy if the ensuing \lp process $\loss=\Theta-\cV $ vanishes almost surely, in which case the ensuing $\mbox{KVA}$ vanishes. As counterparty risk is hard to hedge in practice, the concept of a counterparty risk replicating strategy is a rather theoretical abstraction. A broader notion, which can be related to the notion of optimal replicating portfolio in \citeN{EiseleArtzner11}, would be that of a KVA (or perhaps FTP) minimizing hedge, but even the idea of minimizing over $\eta$ is probably overly optimistic, as banks would not really optimize but rather select a ``suitable'' $\eta.$

The derivative portfolio management strategy of the bank needs to be funded.
We assume that the bank can invest at the OIS rate $r_t$ and obtain unsecured funding at the rate ($r_t+\lambda_t$), where the unsecured funding spread $\lambda$ can be proxied by the bank's CDS spread.
The cash held by the bank, whether borrowed or received as variation margin, is deemed fungible across netting sets in a unique funding set. Initial margin is funded separately at a blended spread $\bar{\lambda}$ depending on the IM funding policy (see \sr{ss:im}).

We assume that \lps (or profits) accumulate before being released at each of an increasing sequence of {$\gg$-predictable} stopping times $0<t_1<\ldots<t_{\nu}  \leq  \tb.$ In other words, the reserve capital process is reset to its theoretical target level $\Theta$ at times $t_l.$ These are typically quarter ends for bank profits, released as dividends, vs recapitalization managerial decision times for losses, and for notational convenience we also introduce $t_0=0$. The random integer $\nu$ corresponds to the last dividend date either prior to default or the one prior to the last maturity in the portfolio. 
At times $t_l$ (for $l\ge 1$), shareholders either realize a loss or a gain $(\loss _{t_l}-\loss _{t_{l -1}}),$
depending on whether reserve capital is below or above its theoretical target value.
If the process $L$ is a continuous-time risk-neutral local martingale, then cumulative realised \lps are given by the discrete time risk-neutral martingale $(\loss _{t_l})_{0\le\nu}$ 
(starting from the accrued loss $\loss_0 =y $),
which expresses the 
viability
of the underlying
pricing rule.

We denote by $\cVb$ the $\Theta$-reset version of the process $\cV,$
such that $\cVb$ is reset to $\Theta$ at each losses realization time $t_l\in (0,\tb]$ and has the same dynamics as the process $\cV$ between these times. Hence, the process $\cVb$ corresponds to the realized reserve capital of the bank, which is the size of the funding pocket provided by reserve capital. 
By definition:
\bl\label{l:sche} For $t\in[0,\tb],$ we have (in two equivalent forms):
\beql{e:ca}
&\cVb_t=\cV_t +\sum_{0<t_l\le t} (\Theta_{t_l}-\Theta_{t_{l-1}}-(\cV_{t_l}-\cV_{t_{l -1}})),\\&
\cV_t=\cVb_t -\sum_{0<t_l\le t} (\Theta_{t_l}-\cVb_{t_l -}),
\eeql
so that
$
\cV=\Theta $ $\iff \cVb = \Theta  
$
(this is the theoretical case of a counterparty risk replicating hedge $\eta$).
In addition, we have
$\cVb_{t_l}=\Theta_{t_l},$ and
$\cVb_{t_l -}=\Theta_{t_{l-1}}+\cV_{t_l}-\cV_{t_{l -1}},$
for every $l=1,\ldots,\nu$. 
\el
 When the frequency of the losses realization schedule $(t_l)$ goes from 0 
(case of a terminal realization of the losses)
to infinity (limiting case of a continuous-time realization of the losses), the realized reserve capital $\cVb$ ``interpolates'' from $\cV$ to  $\Theta$. In the latter limiting case, Lemma \ref{l:sche} does not apply as such, the connection between 
$\cVb$,
RC, and TRC becoming that  \beql{e:cvbt} \cVb=\Theta,\eeql independently of the RC hedge $\eta.$

Note that the retained earnings cashflows $-(  d\mbox{KVA} _t -r_t \mbox{KVA}_t dt )$ also accumulate 
in a separate cash account before being released at the $t_l$, so that the netted dividend at $t_l$ is the negative of
\beql{e:div}
& \loss _{t_l}-\loss _{t_{l -1}}+  \int_{t_{l -1}}^{t_l}(  d\mbox{KVA} _t -r_t \mbox{KVA}_t dt ).  
\eeql 
In case the amount \qr{e:div} is positive, we assume that managers recapitalise the bank (thus diluting existing shareholders) to make up for the shortfall in reserve capital.

In addition to the funding pocket provided by $\cVb$, a bank can also post economic capital as variation margin. To model this circumstance, we imagine having an account for economic capital and a separate one for reserve capital. Economic capital is internally lent to the reserve capital account and is remunerated at OIS rates. The portion of economic capital which is then posted as variation margin is remunerated at OIS rates by the receiving counterparty, while the unused difference is deposited in an external account, thus earning OIS.


\section{Exposures at Defaults}
\label{s:liqco}

In the remainder of the paper we apply the above XVA principles in the concrete setup of a bank engaged into bilateral trading 
of a derivative portfolio 
split into several 
netting sets corresponding to counterparties indexed by $i=1,\ldots, n.$ 

Let $\mbox{MtM}^i_t$ be the mark-to-market of the $i$-th netting set,
i.e.~the trade additive risk-neutral conditional expectation of future discounted promised cash flows, ignoring counterparty risk and funding costs. Let 
\begin{equation}
P^i_t = \mbox{MtM}^i_t - {\rm VM}^i_t
\end{equation}
be the net spot exposure of the $i$-th netting set, i.e.~the valuation of the netting set minus the corresponding (algebraic) variation margin ${\rm VM}^i_t$ received by the bank. 
In addition to the variation margin ${\rm VM}^i_t$ that flows between them, the counterparty $i$ and the bank post respective initial margins $\M^i$ and $\overline{\M}^i_t$ in some segregated accounts.
Finally, we denote by $R_i$ the recovery rate of the counterparty $i$.

In principle, one should explicitly model a positive liquidation period, usually estimated to a few days, corresponding to the time interval between the default of a counterparty (or the bank) and the liquidation of its portfolio.
This liquidation period is important in practice since, once a position is fully collateralized in terms of variation margin, the gap risk related to the slippage of MtM$^i_t$ and to unpaid cash-flows during the liquidation period becomes the first order residual risk and the motivation for the initial margins. 
A positive liquidation period is explicitly introduced in \citeN{Crepey14} and \citeN{CrepeySong15} (see also \citeN{PallaviciniBrigo13bprel}) and involves introducing the random variable:
\begin{equation}\label{e:gapri}
\mbox{MtM}^i_{\tau_i + \delta t}  + \delta \mbox{MtM}^i_{\tau_i + \delta t}  - {\rm VM}^{i}_{\tau_i} 
\end{equation}
where $\delta t$ is the liquidation period and $\delta \mbox{MtM}^i_{\tau_i + \delta t}$ is the accrued value of all the cash flows owed by the counterparty during the liquidation period. 

To alleviate the notation in this paper, we take the limit as $\delta t \to 0$ and approximate \qr{e:gapri} through the value at $\tau_i$ of some $\gg$-optional process
$
Q^i.
$
Such approximation 
is commonly used in numerical simulations for rendering gap risk. 

Another related issue is wrong-way risk, i.e.~the adverse dependence between the XVA exposure of the bank and the credit risk of its counterparties, which can be rendered by a feedback impact of  defaults on the processes MtM$^i$ of the survivors. This impact can also be captured in our variable $Q^i_{\tau_i}$ (see \citeN{CrepeySong15}).

\section{Risk-Neutral XVA FBSDE}
\label{ss:csaxvaeq}

Let RC and $\Theta$ denote putative, to-be-determined reserve capital and target 
reserve capital processes of the bank.

Note that regulators require banks to deduct from CET1 the unilateral UCVA, while the fair valuation of first-to-default risk is given by the metric FTDCVA. This spoils the financial meaning of CET1 as being the fair valuation of equity capital. In this paper, we choose to restore this meaning by postulating that, at the default time of the bank $\tau$, a fictitious wealth transfer of amount 
\beql{e:ucvatau}
\mbox{{UCVA}}_\tau= \E_\tau
\sum_{\tau<\tau_i }\beta_{\tau}^{-1}\beta_{ \tau_i } (1-R_i) (Q^i_{ \tau_i } - \M^{i}_{ \tau_i})^+,
\eeql
occurs from the shareholders to the creditors of the bank. 

In order to emphasize the dependence of $\mbox{EC}$ in $\loss$, we write $\mbox{EC}=\mbox{EC}_t(\loss).$ We write $\delta_t$ for a Dirac measure at time $t.$
Collecting all the shareholders sensitive cash flows in a contra-liability myopic perspective, we obtain:
\beql{jeq:selffinoconsbis}
& {\mbox{$\cV_0=\Theta_0 {-y}$ and,
for $t\in (0 ,\tb],$}} 
\\
& d\cV_t =-
\underbrace{\sum_{i\neq 0}
(1-R_i) (Q^i_{ \tau_i } - \M^{i}_{ \tau_i})^+ \delta_{\tau_i}(dt)}_{\mbox{Counterparty default losses}}\\&
-
\underbrace{\Big((r_t+ \lambda_t) \big(\sum_{i\neq 0} J^i_t  P^i_t  -{\mbox{EC}}_t(\loss) -\cVb_t\big)^+ - r_t \big(\sum_{i\neq 0} J^i_t  P^i_t -{\mbox{EC}}_t(\loss) -\cVb_t\big)^- \Big)dt }_{\parbox{30em}{Costs/benefits of funding the VM of the back-to-back hedge of the portfolio net of the VM of the portfolio itself (this is all included in the $P^i_t$) and of the EC and $\cVb$ funding pockets (see \sr{s:pomapa})}}\\&-\underbrace{(r_t + \bar{\lambda}_t ) \sum_{i\neq 0} J^i_t  \overline{\M}^i_t dt}_{\mbox{$\M$ funding costs}}
+
\underbrace{r_t \big(  \sum_{i\neq 0} J^i_t ( P^i_t+\overline{\M}^i_t)\big) dt}_{\mbox{OIS-earning exchanged VMs
and posted IMs}}\\&-
\underbrace{r_t  {\mbox{EC}}_t(\loss)  dt}_{\mbox{Interest paid on economic capital (see the last paragraph of \sr{s:pomapa})}}
\\& + \underbrace{\eta_t  d\mathcal{M}_t }_{\mbox{Counterparty risk hedging gain}} -
\underbrace{\mbox{UCVA}_{\tau}\delta_{\tau}(dt)}_{\mbox{UCVA transfer at own default time of the bank}} 
\\&=    -
 \sum_{i\neq 0}
(1-R_i) (Q^i_{ \tau_i } - \M^{i}_{ \tau_i})^+ \delta_{\tau_i}(dt) -
\mbox{UCVA}_{\tau}\delta_{\tau}(dt)\\& \qqq- \Big(\lambda_t \big(\sum_{i\neq 0} J^i_t P^i_t -{\mbox{EC}}_t(\loss) -\cVb_t\big)^+ +\bar{\lambda}_t \sum_{i\neq 0} J^i_t  \overline{\M}^i_t  -r_t \cVb_t \Big) dt
+  \eta_t  d\mathcal{M}_t  ,
\label{dRC}
\eeql
where
$\cVb$ is the $\Theta$-reset version of the process $\cV$ as of \qr{e:ca},
which corresponds to the realized reserve capital of the bank.
In view of obtaining a local martingale for the difference $L$ between TRC and RC, 
the forward SDE \qr{jeq:selffinoconsbis} for RC suggests to set
\beql{e:cva-fva}&\Theta_t= \E_t
\sum_{t<\tau_i \leq \tb} (1-R_i) (Q^i_{ \tau_i } - \M^{i}_{ \tau_i})^+
+\E_t \big[ \indi{t<\tau < T} \mbox{UCVA}_{\tau}\big]\\& 
+ \E_t \int_t^{\tb} \Big(\lambda_s\big(\sum_{i\neq 0}J^i_s  P^i_s -{\mbox{EC}}_s(\loss)-\cVb_s\big)^+
+\bar{\lambda}_s  \sum_{i\neq 0}J^i_s  \overline{\M}^i_s
 -r_s  \cVb_s 
 \Big) ds
\sp\ttd.
\label{TRC}
\eeql
In fact, assuming that a pair of processes $(\cV,\Theta)$
satisfies \qr{jeq:selffinoconsbis} and \qr{e:cva-fva},
then we have a continuous-time martingale 
$\loss=\Theta-\cV,$
as desired. 

This leads us to the following formal definition, stated as a risk-neutral local martingale condition in order to avoid unnecessary integrability conditions.
\begin{defi}
{\rm \label{d:h}
Given a {counterparty risk} hedge $\eta,$ let a pair-process $(\cV ,\Theta)$
satisfy
the following bank's risk-neutral XVA FBSDE on $[0,\tb]$:
\begin{equation} \label{jeq:selffinoconsq}
{
\bal
&\Theta_\tb =0 \mbox{ and }  {\loss =\Theta -   \cV} \mbox{ is a risk-neutral local martingale null at time 0,}\\
& \mbox{where $\cV_0={\Theta} _0 -y $ and, for $t\in (0,\tb] $,}
\\&
d\cV_t=-\sum_{i\neq 0}
(1-R_i) (Q^i_{ \tau_i } - \M^{i}_{ \tau_i})^+ \delta_{\tau_i}(dt) -
\mbox{UCVA}_{\tau}\delta_{\tau}(dt)
\\& \qqq- \Big(\lambda_t \big(\sum_{i\neq 0} J^i_t  P^i_t  -{\mbox{EC}}_t(\loss) -\cVb_t\big)^+
+\bar{\lambda}_t \sum_{i\neq 0} J^i_t  \overline{\M}^i_t
-r_t \cVb_t \Big) dt
+  \eta_t  d\mathcal{M}_t , 
\eal
}\end{equation}
in which $\cVb$ is the TRC-reset version of the $\cV$ process (cf.~\qr{e:ca}).
We then call $\Theta$ the bank target reserve capital (or risk-neutral XVA) process
associated with the counterparty risk hedge $\eta,$
with \lp process
$\loss.$ }
\end{defi}



\subsection{Modelling Choice for the Losses Realization Schedule}\label{ss:lim}

The impact of the losses realization schedule $(t_l)$ is quite moderate in the risk-neutral XVA FBSDE \qr{jeq:selffinoconsq}, as it only changes the frequency of the resets of $\cVb.$
This is consistent with the fact that, were it not for the market imperfections considered in this paper,
there would be no impact at all of the schedule of realization of the losses, by the 
\citeN{MM1958} theorem.

In particular, in the limiting case of a continuous-time realization of the losses, with
$\cVb=\Theta$ 
 (cf.~\qr{e:cvbt} and the comments following Lemma \ref{l:sche}),
the TRC equation \qr{e:cva-fva} is rewritten as
\beql{e:cva-fvaprelbis}&\Theta_t= \E_t
\sum_{t<\tau_i \leq \tb} (1-R_i) (Q^i_{ \tau_i } - \M^{i}_{ \tau_i})^+
+\E_t \big[ \indi{t<\tau < T} \mbox{UCVA}_{\tau}\big]\\& 
+ \E_t \int_t^{\tb} \big(\lambda_s (\sum_{i\neq 0}J^i_s  P^i_s -{\mbox{EC}}_s(\loss)-\Theta_s )^+
+\bar{\lambda}_s  \sum_{i\neq 0}J^i_s  \overline{\M}^i_s
 -r_s {\Theta_s}
 \big) ds
\sp\ttd ,
\label{TRC2}
\eeql
i.e.
\beql{e:cva-fvabis}
&\Theta_t=\underbrace{\E_t
\sum_{t<\tau_i  }\beta_{t}^{-1}\beta_{ \tau_i } (1-R_i) (Q^i_{ \tau_i } - \M^{i}_{ \tau_i})^+}_{{ \mbox{{UCVA}}_t}}
+\underbrace{\E_t\int_t^{\tb} \beta_{t}^{-1} \beta_{s} \bar{\lambda}_s  \sum_{i\neq 0} J^i_s  \overline{\M}^i_s   ds }_{{ \mbox{{MVA}}_t}}
\\&+\underbrace{\E_t\int_t^{\tb} \beta_{t}^{-1} \beta_{s} \lambda_s \Big(\sum_{i\neq 0} J^i_s  P^i_s -{\mbox{EC}}_s(\loss) -
{\Theta_s}
\Big)^+  ds }_{{ \mbox{{FVA}}_t}} \sp 0\le t\le \tb
,\eeql
which is consistent with the balance sheet equation \qr{e:bsheet-entry-incr-delta-post} whilst giving a more precise meaning to the UCVA, FVA and MVA terms.
Note that the {UCVA} and {MVA} processes in \qr{e:cva-fvabis} are exogenous,
so that \qr{e:cva-fvabis} is equivalent to defining
$\Theta=\mbox{{UCVA}}+ \mbox{{MVA}}+
 \mbox{{FVA}}$  where
FVA is given implicitly through
\beql{e:cva-fvafva}
&\mbox{{FVA}}_t=   \E_t\int_t^{\tb} \beta_{t}^{-1} \beta_{s} \lambda_s \Big(\sum_{i\neq 0} J^i_s  P^i_s-{\mbox{EC}}_s(\loss) \\&\qqq\qqq -
 \mbox{{UCVA}}_s -
 \mbox{{MVA}}_s -
 \mbox{{FVA}}_s 
\Big)^+  ds   \sp 0\le t\le \tb
.\eeql

By comparison, in the other extreme case $\cVb=\cV$ case of 
a terminal realization of the losses,
the TRC equation \qr{e:cva-fva} that is implicit in the martingale condition
for $L$ in \qr{jeq:selffinoconsq}
is rewritten as (assuming integrability)
\beql{e:cva-fvaprel}&\Theta_t= \E_t
\sum_{t<\tau_i \leq \tb} (1-R_i) (Q^i_{ \tau_i } - \M^{i}_{ \tau_i})^+
+\E_t \big[ \indi{t<\tau < T} \mbox{UCVA}_{\tau}\big]\\& 
+ \E_t \int_t^{\tb} \Big(\lambda_s\big(\sum_{i\neq 0}J^i_s  P^i_s -{\mbox{EC}}_s(\loss)-\cV_s\big)^+
+\bar{\lambda}_s  \sum_{i\neq 0}J^i_s  \overline{\M}^i_s
 -r_s {\cV_s}
 \Big) ds
\sp\ttd
\label{TRC1}
\eeql
The connection with the balance-sheet equation
\qr{e:bsheet-entry-incr-delta-post} is a bit less nice
than with \qr{e:cva-fvabis}.

As a consequence, since a modelling choice has to be made regarding the $(t_l)$ anyway,
we assume henceforth a continuous-time realization of the losses, with
$\cVb=\Theta$ in \qr{jeq:selffinoconsq}, or, equivalently (assuming integrability),
a TRC as of \qr{e:cva-fvabis} and a FVA as of \qr{e:cva-fvafva}.

\subsection{Entry Price Versus Fair Valuation}\label{rem:dva}
The TRC of Definition
\ref{d:h}
is only fair to the shareholders of the bank and if the cost of capital is ignored. As already seen from a balance sheet point of view in \sr{ss:bs},
an approach fair to the bank as a whole (shareholders plus bondholders altogether) and to its counterparties,
i.e.~a symmetric XVA approach, reckons contra-liabilities: 
\begin{itemize}\item The DVA of the bank, which  correspond to the CVA from the point of view of its counterparties; \item The difference between the UCVA and the FTDCVA;
\item The 
FDA (akin to the DVA2 in \citeN{HullWhite13d}), which is equal to the FVA;  
\item The MDA, which is equal to the MVA.
\end{itemize}
Adding all the contra-liability cash-flows into \qr{jeq:selffinoconsbis} and exploiting the above-mentioned cancellations (cf.~also \sr{ss:bs}), we would end-up with the following alternative to \qr{e:cva-fvabis} (in the limiting case \qr{e:cvbt} of a continuous-time realization of the losses):
\beql{e:cva-fvater}
&\Theta_t=\underbrace{\E_t
\sum_{t<\tau_i \le \tb }\beta_{t}^{-1}\beta_{ \tau_i } (1-R_i) (Q^i_{ \tau_i } - \M^{i}_{ \tau_i})^+}_{{ \mbox{{FTDCVA}}_t}}
\\&\qqq\qqq-\underbrace{\sum_{t<\tau_i \le \tb }\beta_{t}^{-1}\beta_{ \tau_i } (1-R_i) (Q^i_{ \tau_i } - \overline{\M}^{i}_{ \tau_i})^-}_{{ \mbox{{FTDDVA}}_t}} \sp 0\le t\le \tb
.\eeql 
Such a symmetric TRC yields an objective, reference metric for the FTP, ignoring the market inefficiencies accounted for by the entry price XVA given as 
$
\mbox{UCVA}+\mbox{FVA}+\mbox{MVA}+\mbox{KVA} $
(cf.~\qr{e:cva-fvabis} and \qr{e:bsheet-entry-incr-delta}).

\subsection{About Initial Margins}\label{ss:im}

The MVA
depends strongly on the strategy that is postulated regarding the funding of the initial margins. 
For instance, 
instead of an unsecured funding scheme resulting in $\bar{\lambda}=\lambda,$
one can assume that initial margins are funded through a specialist lender that lends only IM and, in case of default of the bank, receives back the portion of IM unused to cover losses. The
exposure of the specialist lender to the default of the bank is given by 
$  (1-R)  \sum_{i\neq 0}J^i_{\tau} \big(( Q^i_{\tau})^- \wedge \overline{\M}^i_{\tau}\big),$ 
where $R$ is the bank recovery rate (typically taken as 40\%).
Denoting by $\gamma$ the risk-neutral default intensity process of the bank,
the ensuing instantaneous IM funding charge for the bank is
(assuming here for simplicity $\gg$-predictable processes $Q^i$)
$$\gamma_t (1-R)\sum_{i\neq 0}J^i_{t} \big(( Q^i_{t})^- \wedge \overline{\M}^i_{t}\big)= {\lambda_t}\sum_{i\neq 0}J^i_{t} \big(( Q^i_{t})^- \wedge \overline{\M}^i_{t}\big).$$
By identification with the general form $\bar{\lambda}_t  \sum_{i\neq 0}J^i_{t}   \overline{\M}^i_{t}$ postulated for instantaneous initial margin costs in \qr{jeq:selffinoconsbis}, this corresponds to
$$\bar{\lambda}_t=\frac{\sum_{i\neq 0}J^i_{t} \big(( Q^i_{t})^- \wedge \overline{\M}^i_{t}\big)}{\sum_{i\neq 0}J^i_{t}   \overline{\M}^i_{t}}   {\lambda}_t.$$
Such a blended spread $\bar{\lambda}_t$ can be much smaller than the unsecured funding spread $\lambda$.
A more detailed discussion 
is in \shortciteN{AAI2015}.

\part{Implementation}\label{s:impl}

In this final part of the paper we implement the above XVA approach by means of nested Monte Carlo simulations that are used for solving
the risk-neutral XVA FBSDE \qr{jeq:selffinoconsq} for (RC,TRC)
iteratively, in the limiting 
case \qr{e:cva-fvabis} of a continuous-time realization of the losses (i.e.~$\cVb=\Theta$). The ensuing loss process  $L=\Theta-\cV$ is then plugged as input data of the KVA computations. The paper is concluded by two case studies.

\section{Iterative Algorithm}\label{s:iter}

\subsection{Counterparty Risk Replication BSDE}\label{ss:csaxvarepl}

Unless $\lambda=0,$ the FBSDE \qr{jeq:selffinoconsq}
 is made nonstandard by the term ${\mbox{EC}}_t(\loss),$ which entails the conditional law of the one-year-ahead increments of the process $\loss=\Theta-\cV$. As a starting point in the search for a solution to our FBSDE, we define
the following counterparty risk replication BSDE for a process $\Theta^\star:$
\beql{e:bvabsde}
&\Theta^\star_t=\underbrace{\E_t
\sum_{t<\tau_i  }\beta_{t}^{-1}\beta_{ \tau_i } (1-R_i) (Q^i_{ \tau_i } - \M^{i}_{ \tau_i})^+}_{{ \mbox{{UCVA}}_t}}
+\underbrace{\E_t\int_t^{\tb} \beta_{t}^{-1} \beta_{s} \bar{\lambda}_s  \sum_{i\neq 0} J^i_s  \overline{\M}^i_s   ds }_{{ \mbox{{MVA}}_t}}
\\&\qqq +\underbrace{\E_t\int_t^{\tb} \beta_{t}^{-1} \beta_{s} \lambda_s \Big(\sum_{i\neq 0} J^i_s  P^i_s  -\Theta^\star_s\Big)^+  ds }_{{ \mbox{{FVA}}^\star_t}} \sp 0\le t\le \tb
.\eeql

Recall that $\cH^2$ denotes the space of ${\cdot}^2$-integrable processes over $[0,\tb].$

\bt Assuming that $r$ and $\lambda$ are bounded from below and
that $r$ and \\
$\lambda  (\sum_{i\neq 0}J^i   P^i   -\mbox{UCVA}   -\mbox{MVA} ) $ are 
in $\cH^2$, then the counterparty risk replication BSDE \qr{e:bvabsde} has a unique solution $\Theta^\star$ in $\cH^2$ .
\et
\proof
Noting that the ($\mbox{UCVA}+\mbox{MVA}$) process is exogenous, the BSDE \qr{e:bvabsde} is
equivalent to the following BSDE for the FVA$^\star$ process:
\beql{e:fvabsde}
\beta_t \mbox{FVA}^\star_t=\E_t \int_t^{\tb} \beta_s\lambda_s \Big(\sum_{i\neq 0}J^i_s  P^i_s  -\mbox{UCVA}_s  -\mbox{MVA}_s - \mbox{FVA}^\star_s\Big)^+   ds  \sp 0\le t\le \tb,
\eeql
which is well-posed in $\cH^2$ under the assumptions of the theorem,
 by the monotonic generator BSDE
arguments already used in the proof of Lemma \ref{l:wp}.~\finproof\\
 
\noindent
If $\eta  d\cM $ matches the martingale part of $\Theta^\star$ for some hedge $\eta=\eta^\star,$
then we can readily verify that the pair process $(\cV,\Theta)=(\Theta^\star,\Theta^\star)$ yields a solution to the risk-neutral XVA FBSDE \qr{jeq:selffinoconsq} associated with $\eta=\eta^\star.$ Since the resulting process $\loss$ vanishes, so do the corresponding EC and KVA processes
and the ensuing FVA is given by $\mbox{FVA}^\star.$

\subsection{Main Loop}\label{ss:csaxvadisc}

In practice,
the depth of the counterparty risk hedging market is very far from the replicating point where $\eta=\eta^\star$ and $\loss=0$ (not to say opposite to it, i.e.~$\eta=0$).
Consequently, unless $\lambda=0$ (which may be realistic for some insurance cases but rarely so for banks),
the replication XVA $\Theta^\star$ should only be seen as the starting point for
the following
iteration in the search for a solution to the FBSDE \qr{jeq:selffinoconsq} (in the $\cVb=\Theta$ case
of a continuous-time realization of the losses):
$\cV^{(0)}=\Theta^{(0)}-y=\Theta^\star -y$ and, for $k\ge 1,$
\begin{equation} \label{e:prat-csa}
{
\bal
&\cV^{(k)}_0=\Theta^{(k-1)}_0 -y \mbox{ and, for $t\in (0 ,\tb]$,}
\\
&d\cV^{(k)}_t =  r_t\Theta^{(k-1)}_t dt
-\sum_{i\neq 0} (1-R_i) (Q^i_{ \tau_i } - \M^{i}_{ \tau_i})^+ \delta_{\tau_i}(dt) -
\mbox{UCVA}_{\tau}\delta_{\tau}(dt)
\\&  - \lambda_t\Big(\sum_{i\neq 0} J^i_t  P^i_t
 -{\mbox{EC}}_t \big(\Theta^{(k-1)}-\cV^{(k-1)}    \big) 
- {\Theta^{(k-1)}_{t}}
\Big)^+   dt
\\&- \bar{\lambda}_t \sum_{i\neq 0} J^i_t  \overline{\M}^i_t dt+
\eta_t  d\mathcal{M}_t ;
 \\
& \Theta^{(k)}_t= \E_t
\sum_{t<\tau_i \leq \tb}\beta_t^{-1} \beta_{\tau_i} (1-R_i) (Q^i_{ \tau_i } - \M^{i}_{ \tau_i})^+
+\E_t \big[\beta_t^{-1}\beta_{\tau }  \indi{t<\tau < T} \mbox{UCVA}_{\tau}\big]
\\&
+ \E_t \int_t^{\tb}\beta_t^{-1}\beta_{s}  \lambda_s\big(\sum_{i\neq 0}J^i_s  P^i_s -{\mbox{EC}}_t \big(\Theta^{(k-1)}-\cV^{(k)}
\big)-
{\Theta^{(k-1)}_{s}}
\big)^+ ds
\\&+ \E_t \int_t^{\tb}\beta_t^{-1}\beta_{s}   \bar{\lambda}_s  \sum_{i\neq 0} J^i_s  \overline{\M}^i_s   ds\sp\ttd.
\eal
}\end{equation}



However, going through this loop
numerically
necessitates a dynamic and iterative simulation of economic capital processes $\mbox{EC}(\loss)=\max(\RM(\loss) ,\mbox{KVA}(\loss)),$ including conditional risk measure simulations and \mbox{KVA} BSDE solutions based on these. Some approximations are required for the sake of tractability 
(especially on real life large portfolios as of \sr{ss:lp}).

In practice, we suggest to pass only once in the loop \qr{e:prat-csa}, using $\Theta^{(1)}$ as risk-neutral XVA and
$\loss^\star=\Theta^\star-\cV^{(1)}$ to compute $\RM(\loss^\star)$
 and the ensuing {KVA}$(\loss^\star)$ 
via the KVA BSDE \qr{e:kva}. A slightly simpler alternative is to use
the
linear KVA formula
\qr{e:kexpl} based on $C=\RM(\loss^\star),$ checking that the ensuing KVA is lower than $\RM(\loss^\star)$ (see the remark \ref{r:ecr}).
In addition, 
we  use a deterministic term structure 
\beql{e:tsa}{\mbox{ES}}_\star(t)\approx {\mbox{ES}_t} (\loss^\star)  
\eeql
{obtained by projecting in time instead of conditioning with respect to $\cG_t$.  

From a theoretical perspective,
controlling the iteration \qr{e:prat-csa} for establishing its convergence to 
\qr{jeq:selffinoconsq}
is challenging due to the terms of the form ${\mbox{EC}}_t (\loss)$, which involve in a very nonlinear fashion the conditional law of the one-year-ahead increment of $\loss.$
The mathematical study 
of the well-posedness of the exact FBSDE \qr{jeq:selffinoconsq} 
and of the convergence toward it of the iterative scheme
\qr{e:prat-csa}, of no direct use in this work,
are deferred to a separate BSDE paper.

\section{Case Studies}
\label{sec:case-study}

As case studies, we present XVA computations on foreign-exchange and fixed-income portfolios. Toward this end we use the market risk and portfolio credit risk models of \citeN{AlbaneseBellajGimonetPietronero11} 
calibrated to the relevant market data. We assume no counterparty risk hedge (i.e.~$\eta=0$) and no
margins on the portfolio, i.e.~no variation or initial margins (but perfect variation-margining on the portfolio back-to-back hedges). In particular, 
the $\mbox{MVA}$ numbers are all equal to zero and hence not reported in the tables below. 

All the computations are run using a 4-socket server for Monte Carlo simulations, NVIDIA GPUs for algebraic calculations and Global Valuation Esther as simulation software.
We use nested simulation with primary scenarios and secondary scenarios generated under the risk neutral measure calibrated to derivative data using broker datasets for derivative market data.    
For the purpose of the expected shortfall computations we use a historical measure on the secondary scenarios obtained by adjusting the drifts of the risk-neutral model in such a way to satisfy backtesting benchmarks.


Based on the nested simulated paths, we first solve the counterparty risk replication BSDE \qr{e:bvabsde} for the process $\mbox{TRC}^\star.$
We also obtain UCVA$_0$ and a preliminary FVA value as $\mbox{FVA} ^\star_0=\mbox{TRC}^\star _0- \mbox{UCVA} _0$.

The process $\cV^{(1)}$ is then simulated based on the related forward SDE  with $\eta=0$ in \qr{e:prat-csa} (note the EC term there vanishes in case $k=1$).

The paths of $\loss^\star=\Theta^\star - \cV^{(1)}$ are  used for inferring a term structure ${\mbox{ES}}_\star(t)\approx {\mbox{ES}_t} (\loss^\star)$
as of \qr{e:tsa}. 
This term structure
is plugged as $C$
in the linear formula \qr{e:kexpl} for computing the KVA (see the last paragraph of \sr{ss:csaxvadisc}), assuming a hurdle rate of 10.5\%.

The backward SDE for $\Theta^{(1)}$ in \qr{e:prat-csa} is then used
with the above approximation ${\mbox{ES}}_\star$ for the $\mbox{EC}$ term in order to obtain a refined FVA value as $\mbox{FVA}_0 \approx \Theta^{(1)}_0-\mbox{UCVA}_0.$

\subsection{Toy Portfolio}\label{ss:te}

We first consider a portfolio of
ten USD currency swaps depicted in Table \ref{t:swaps},
on the date of 11 January 2016. 
The nominal of each swap is $10^4$.
The swaps are traded with four counterparties, with 40\% recovery rate 
and credit curves (also of the bank) as of Table \ref{t:cds}.

We use 20,000 primary scenarios up to 30 years in the future run on 54 underlying time points
with 1,000 secondary scenarios starting at each of these, which amounts to a total of $20,000\times 54\times 1,000=$ 1,080 million scenarios. 
The whole calculation takes roughly 10 minutes to run (including the nested simulations for capital).
The corresponding XVA results are displayed in Table \ref{t:xvas}.
Since the portfolio is not collateralized, its UCVA is quite high. But its KVA
is even higher. The FVA is much smaller as it is eroded by the collateral pockets provided by economic capital (and also, to some lesser extent, by realized reserve capital).
The reference ``fair'' FTDCVA and FTDDVA metrics are also shown for comparison. Note that given our deterministic term structure approximation \qr{e:tsa} for expected shortfalls, the computation of the KVA based on it reduces to a deterministic time-integral (which explains why there is no related standard deviation error in Table \ref{t:xvas}).

Table \ref{t:incrxvas} shows the incremental XVA results when Swap 5 (resp.~9)  is last added in the portfolio. Interestingly enough, all the incremental XVAs of Swap 9 (and also the incremental FVA of Swap 5) are negative. Hence, Swap 9 is XVA profitable to the portfolio, meaning that a price maker should be ready to enter the swap for less than its mark-to-market value, assuming it is already trading the rest of the portfolio. This result illustrates the importance of the endowment in modern derivative portfolio management.
\begin{table}[htbp]
\begin{center}\begin{tabular}{|r|r|r|r|r|r|}
\hline
Type & ID & Maturity & Receiver Rate & Payer Rate & Netting Set\\
\hline
Swap & 1 & 10y & Par – 6M & LIBOR – 3M & C\\
\hline
Swap & 2 & 10y & LIBOR – 3M & Par – 6M & B\\
\hline
Swap & 3 & 5y & Par – 6M & LIBOR – 3M & B\\
\hline
Swap & 4 & 5y & LIBOR – 3M & Par – 6M & C\\
\hline
Swap & 5 & 30y & Par – 6M & LIBOR – 3M & B\\
\hline
Swap & 6 & 30y & LIBOR – 3M & Par – 6M & A\\
\hline
Swap & 7 & 2y & Par – 6M & LIBOR – 3M & A\\
\hline
Swap & 8 & 2y & LIBOR – 3M & Par – 6M & D\\
\hline
Swap & 9 & 15y & Par – 6M & LIBOR – 3M & A\\
\hline
Swap & 10 & 15y & LIBOR – 3M & Par – 6M & D\\
\hline
\end{tabular}\end{center}
\caption{Portfolio of USD currency swaps.}\label{t:swaps}
\end{table} 
\begin{table}[htbp]
\begin{center}\begin{tabular}{|r|r|r|r|r|r|r|r|r|r|r|r|r|}
\hline
& 6M & 1Y & 2Y & 3Y & 4Y & 5Y & 7Y & 10Y & 15Y & 20Y & 30Y & 50Y\\
\hline
A & 9 & 11 & 18 & 25 & 33 & 41 & 55 & 65 & 68 & 70 & 71 & 70\\
\hline
B & 12 & 15 & 25 & 37 & 49 & 62 & 79 & 91 & 95 & 97 & 98 & 98\\
\hline
C & 44 & 59 & 102 & 146 & 191 & 235 & 276 & 295 & 301 & 302 & 304 & 304\\
\hline
D & 272 & 291 & 327 & 367 & 400 & 433 & 452 & 460 & 460 & 462 & 461 & 461\\
\hline
Bank & 18 & 24 & 40 & 58 & 77 & 96 & 120 & 134 & 139 & 141 & 142 & 142\\
\hline
\end{tabular}\end{center}
\caption{Credit curves of the counterparties and the bank.}\label{t:cds}
\end{table}
\begin{table}[htbp] 
\begin{center}\begin{tabular}{|r|r|r|}
\hline
& \$Value & Standard Rel. Error\\
\hline
FTDCVA$_0$ & 372.22 & 0.46\%\\
\hline
FTDDVA$_0$ & 335.94 & 0.51\%\\
\hline
UCVA$_0$ & 471.23 & 0.46\%\\
\hline
FVA$_0$ & 73.87 & 1.06\%\\
\hline
KVA$_0$ & 668.83 & {N/A}\\
\hline
\end{tabular}\end{center}
\caption{XVA values for the toy portfolio.} \label{t:xvas}
\end{table}
\begin{table}[htbp] 
\begin{center}\begin{tabular}{|r|r|}
\hline
\multicolumn{2}{|c|}{Incremental \$Value for Swap 5}\\
\hline
FTDCVA$_0$ & 98.49\\
\hline
FTDDVA$_0$ & 122.91\\
\hline
UCVA$_0$ & 155.46
\\
\hline
FVA$_0$ & -85.28\\
\hline
KVA$_0$ & 127.54\\
\hline
\end{tabular}\begin{tabular}{|r|r|}
\hline
\multicolumn{2}{|c|}{Incremental \$Value for Swap 9}\\
\hline
FTDCVA$_0$ & -23.83\\
\hline
FTDDVA$_0$ & -73.63\\
\hline
UCVA$_0$ & -27.17
\\
\hline
FVA$_0$ & -8.81\\
\hline
KVA$_0$ & -52.85\\
\hline
\end{tabular}\end{center}
\caption{Incremental XVAs when a swap is added last into the portfolio. {\it Left}: Impact of Swap 5. {\it Right}: Impact of Swap 9. } \label{t:incrxvas}
\end{table}

\subsection{Large Portfolio}\label{ss:lp}

We now consider a representative fixed-income portfolio with about 2,000 counterparties, 100,000 fixed income trades including swaps, swaptions, FX options, inflation swaps and CDS trades.

\begin{table}[h]
\centering 
\begin{tabular}{|r |r| r| } 
\hline
 XVA & \$Value 
\\  
\hline 
UCVA$_0$ & 242 M   \\ 
\hline
FVA$_0^\star$ & 126 M \\ 
\hline
FVA$_0$   & 62 M \\ 
\hline
KVA$_0$ & 275 M \\ 
\hline 
\end{tabular}
\caption{XVA values for the large portfolio.} 
\label{tab:XVA-results}
\end{table}

We use 20,000 primary scenarios up to 50 years in the future run on 100 underlying time points
with 1,000 secondary scenarios starting at each of these, which amounts to a total of two billion scenarios.  In this case the whole calculation takes 3 hours.
Table \ref{tab:XVA-results} shows the XVA results for the case study portfolio.

The KVA amounts to \$275 M, which makes it the greatest of the 
XVA numbers, roughly fifteen percent above the (even though uncollateralized) UCVA.
The left panel in Figure \ref{fig:EC} shows the term structure of economic capital along with the term structure of the KVA obtained by the deterministic term structure
approximation ${\mbox{ES}}_\star$ as of 
 \qr{e:tsa} for economic capital and the linear formula
\qr{e:kexpl} for the KVA. Note that the KVA obtained in this way is at all times below
the corresponding expected shortfall (cf.~the remark \ref{r:ecr}).

We report two FVA metrics to show how the FVA gets reduced when we consider additional funding sources. 
The number FVA$^\star$ accounting only for re-hypothecation of variation margin received on hedges amounts to approximately \$126 M. However, if we consider the additional funding sources due to economic capital and realized reserve capital, we arrive at an FVA figure of \$62 M, less than a half of the re-hypothecation-only FVA. The funding needs' reduction achieved by EC, UCVA and FVA is also shown in the right panel of Figure \ref{fig:EC} by the
FVA blended curve. This is the FVA funding curve which, whenever applied to the FVA
computed neglecting the impact of economic and reserve capital, gives rise to the same term structure
for the forward FVA as the calculation carried out including instead capital in the calculation as a source
for funding. This blended curve is often inferred by consensus estimates based on the Markit XVA service.
However, here it is computed from the ground up based on full-fledged capital projections.

\begin{figure}[h!]
\centering
\includegraphics[width=0.49\textwidth,height=0.25\textheight]{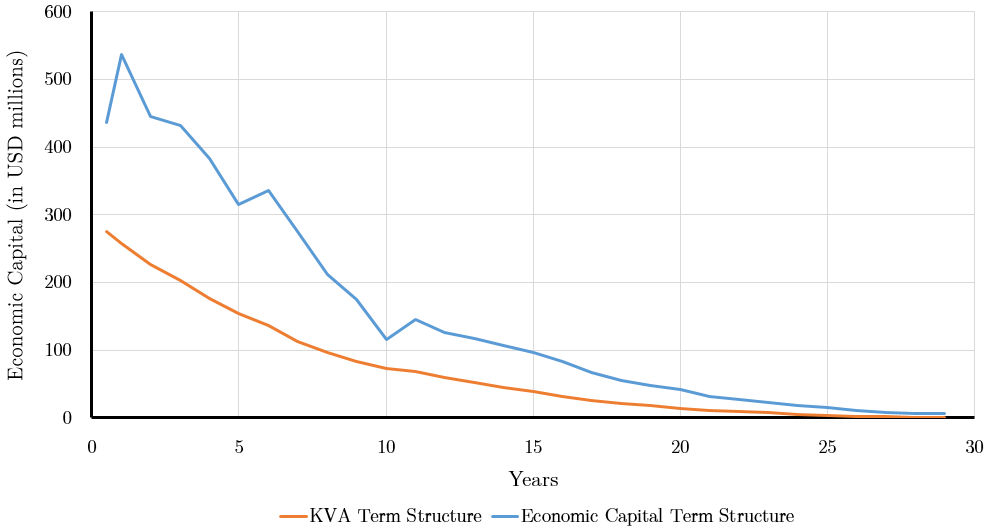}
\includegraphics[width=0.49\textwidth,height=0.27\textheight]{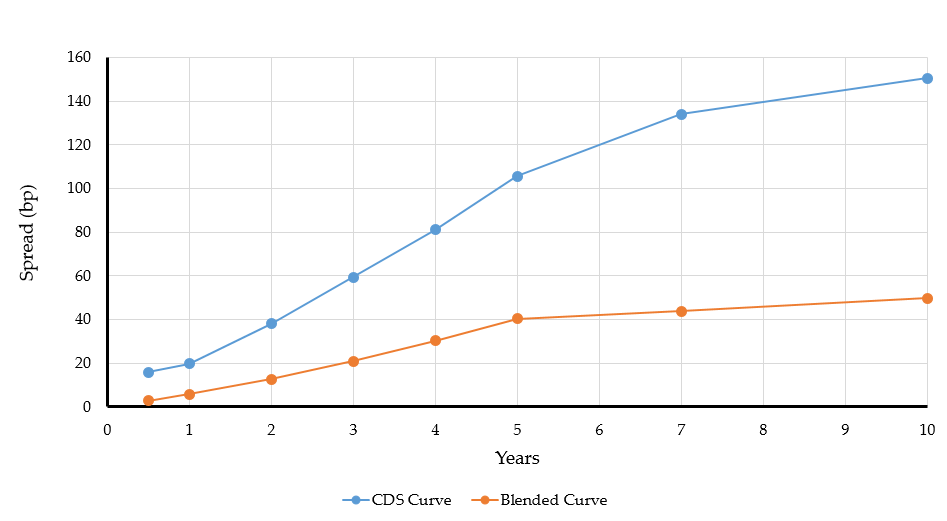}

\caption{{\it (Left)} Term structure of economic capital computed as the expected shortfall with 97.5\% confidence for capital return distribution compared with the term structure of KVA. {\it (Right)} FVA blended curve computed from the ground up based on capital projections.}
\label{fig:EC}
\end{figure}





\appendix

\section{Further Discussions}\label{s:concl}

\subsection{KVA Is Not a CET1 Deduction}\label{ss:notded}

In \citeN{GreenKenyonDennis14} 
 {as also discussed in some theoretical actuarial literature (see \citeN[Sect.~4.4]{SalzmannWuthrich2010})}, the KVA is treated as yet another risk neutral XVA which enters with an additional $(  d\mbox{KVA} _t -r_t \mbox{KVA}_t dt )$ term in $\cV$ equations such as \qr{jeq:selffinoconsbis}.
The term $(  d\mbox{KVA} _t -r_t \mbox{KVA}_t dt )$ is treated as an additional cash-flow released by the XVA trader to the shareholders, as if, in particular, there was a single account for reserve capital and retained earnings altogether. 

From the balance-sheet point of view of \sr{ss:ep}  (cf.~Figure \ref{fig:bs}), this is tantamount to treating the KVA as a CET1 deduction, i.e.~considering the theoretical target value \TKVA of retained earnings
(e.g. the solution to our KVA BSDE \qr{e:kva})
 as a further contra-asset in \qr{e:bsheet-entry}.
This results in an equity 
defined from the following balance sheet equation (compare with \qr{e:bsheet-entry}):
\bel
\mbox{Assets}-\underbrace{(\mbox{UCVA}+\mbox{FVA}+\mbox{MVA}
+\mbox{\TKVA})}_{\mbox{Contra-assets}}
=\mbox{Liabilities} 
+  \mbox{Equity}.
\eel
One then obtains an
ersatz of the FTP rule
\qr{e:bsheet-entry-incr-delta}
by postulating a flat Equity at a new deal, which leads 
to
\beql{e:bsheet-entry-incr-delta-post-direct}
\mbox{FTP}=\Delta(\mbox{Assets}-\mbox{Liabilities})
=\Delta\mbox{UCVA}+\Delta\mbox{FVA}
+
\Delta\mbox{MVA}+\Delta\mbox{\TKVA}.
\eeql
This is formally the same FTP formula as \qr{e:bsheet-entry-incr-delta},
but the KVA and the FVA components in the two formulas differ.
Namely, in \qr{e:bsheet-entry-incr-delta}, the $\mbox{TRC}=\mbox{UCVA}+ \mbox{FVA}+ \mbox{MVA}$ is computed in a first step by postulating a flat CET1 net of the KVA account (cf.~Figure \ref{fig:bs} and \qr{e:bsheet-entry-incr-delta-prel}). The KVA is then obtained by using the ensuing loss process $\mbox{L}=\mbox{TRC}-\mbox{RC}$ as input data in the KVA BSDE \qr{e:kva}. By contrast, with \qr{e:bsheet-entry-incr-delta-post-direct}, the sum between TRC and \TKVA is computed in one shot by postulating a flat Equity, subject to the constraint that the \TKVA satisfies the KVA BSDE \qr{e:kva} with input data $L=\mbox{TRC}-\mbox{RC}.$ This is not only erroneous conceptually, but untractable numerically unless simplifying approximations are made to decouple the flat Equity equation from the constraint, such as working with regulatory instead of economic capital (otherwise KVA fluctations should be simulated for capital calculations) and ignoring that the KVA is loss-absorbing and in \shortciteN{GreenKenyonDennis14}.

\subsection{About the Solvency Accounting Condition}\label{ss:solv}

As discussed after \qr{e:adm}, our Solvency-like accounting condition
$C\geq K(C)$ is meant to express a switching (or transferability) property of the bank to new shareholders at any time if need be.
Accounting for a losses realization schedule $(t_l)$ and the ensuing reserve capital shortfall
$\lossb= \Theta-\cVb$ (see Lemma \ref{l:sche} and the surrounding comments), a relevant refinement of this condition should be 
$C\geq K(C)+\lossb.$ However, under this refined constraint and for the correspondingly amended form of the set of admissible economic capital processes $\mathcal{C}$ (cf.~\qr{e:adm}),
the mathematical solution of the capital at risk optimization problem in Part \ref{s:coc} becomes less clear. But, 
in the end, we work under the modeling assumption of a continuous-time realization of the losses, in which case 
$\lossb$ vanishes identically. In this case the difference between the basic accounting condition
$C\geq K(C)$ and the refined accounting condition $C\geq K(C)+\lossb$ is immaterial.


\newpage
\section*{Main Acronyms}
\label{sec:acronyms}
 
\begin{description}
\item{\bf CDS} Credit default swap.
\item{\bf CET1} Common Equity Tier I Capital.
\item{\bf CVA} Credit valuation adjustment (can be either UCVA or FTDCVA).
\item{\bf DVA} Debt valuation adjustment  (can be either UDVA or FTDDVA).
\item{\bf DVA2} Funding debt adjustment (same as FDA).
\item{\bf EC} Economic capital.
\item{\bf ES} Expected shortfall.
\item{\bf FDA} Funding liability adjustment (same as DVA2).
\item{\bf FRTB} Fundamental Review of the Trading Book.
\item{\bf FTDCVA} First-to-default CVA.
\item{\bf FTDDVA} First-to-default DVA.
\item{\bf FTP} Funds transfer pricing.
\item{\bf FVA} Funding valuation adjustment.
\item{\bf IFRS} International Financial Reporting Standards. 
\item{\bf IM} Initial margin.
\item{\bf KVA} Capital valuation adjustment (the banking notion of risk margins)
\item{\bf MtM} Mark-to-market.
\item{\bf MVA} Margin valuation adjustment.
\item{\bf OIS rate} Overnight index swap rate.
\item{\bf RACET1} Risk-adjusted CET1.   
\item{\bf RC} Reserve capital.
\item{\bf SCR} Shareholders' capital at risk.
\item{\bf TRC} Target reserve capital (same as risk-neutral XVA).
\item{\bf UCVA} Unilateral CVA. 
\item{\bf UDVA} Unilateral DVA. 
\item{\bf VM} Variation margin.
\item{\bf XVA} Generic ``X'' valuation adjustment 
\end{description}

\end{document}